\documentclass[prb, letter, superscriptaddress]{revtex4}
\usepackage{amssymb}
\usepackage{amsmath}
\usepackage{bbm}
\usepackage{graphicx}

\begin{document}

\author{Philipp Werner}
\affiliation{Theoretische Physik, ETH Zurich, 8093 Zurich, Switzerland}
\author{Takashi Oka}
\affiliation{Department of Physics, Tokyo University, Hongo, Tokyo 113-0033, Japan}
\author{Andrew J. Millis}
\affiliation{Department of Physics, Columbia University, 538 West, 120th Street, New York, NY 10027, USA}

\title{Diagrammatic Monte Carlo simulation of non-equilibrium systems}

\date{December 22, 2008}

\hyphenation{}

\begin{abstract}
We generalize the recently developed diagrammatic Monte Carlo techniques for quantum impurity models from an imaginary time to a Keldysh formalism suitable for real-time and nonequilibrium calculations. Both weak-coupling and strong-coupling based methods are 
introduced, analysed 
and applied to the study of transport and relaxation dynamics in interacting quantum dots. 
\end{abstract}


\maketitle

\section{Introduction}

Quantum impurity models play a prominent role in nanoscience as mathematical representations of quantum dots, single-molecule devices and adatoms on surfaces. In general theoretical terms a quantum impurity model is a  system with a finite-dimensional Hilbert space (``dot") coupled to one or more infinite systems (``baths") described by a Hilbert space with a continuum of energy levels. The equilibrium properties of quantum impurity models are by now reasonably well understood theoretically and indeed in  most cases the properties of interest can be computed numerically to the necessary accuracy. 

By contrast, the nonequilibrium properties of quantum impurity models are much less well understood. The subject is of fundamental theoretical importance, as an instance of the basic problem of the properties of nonequilibrium quantum many-body systems.  It is also of considerable experimental interest in connection with the properties of quantum dots where the Kondo effect plays an important role in transport properties. \cite{Ng88,Glazman88,Goldhaber-Gordon98,Cronenwett98,vanderWiel00} Quantum impurity models are also closely connected to the issue of transition rates and reaction dynamics in chemistry. 

Quantum impurity models may be driven out of equilibrium in several ways. 
If a system is coupled to more than one reservoir, then a chemical potential or temperature difference between reservoirs can generate a nonequilibrium steady state in which current flows from one reservoir to another across the dot. One may also consider a transient or steady-state irradiation of the dot  or the relaxation to steady state of an atypical initial condition; in either case one may have equilibrium or nonequilibrium reservoirs. While the basic formalism for dealing with these problems was established by Schwinger \cite{Kadanoff} and Keldysh \cite{Keldysh64} in the early 1960s,  and a wide variety of perturbative approaches have appeared (mainly tailored to specific physical applications), it is  important to develop unbiased numerical methods which allow to test theoretical conjectures and to compare the properties of theoretical models to phenomena seen in experiments. 

Several numerical techniques have  been applied to time dependent problems in interacting quantum dots. Numerical renormalization group methods \cite{Anders06} have been shown to provide impressively accurate treatments of  relaxation dynamics in dots with equilibrium baths and extensions to nonequilibrium baths have recently been proposed.\cite{Anders08} However, experience in equilibrium problems has been that these approaches, although powerful, are limited in the range of problems that can be treated and the range of energy scales that can be accessed. Path integral sampling techniques introduced in the quantum chemistry context \cite{Makri99} have recently been extended to the quantum dot problem.\cite{Weiss08} These techniques  require a finite ``memory time", and are therefore restricted to non-zero temperature and voltage bias.  The time-dependent non-crossing approximation \cite{Goke07} gives access to long times and spectral functions, but is probably not reliable at strong interactions. The time-dependent density matrix renormalization group was also used to study the transport properties of quantum dots coupled to one-dimensional reservoirs. \cite{Al-Hassanieh06, Kirino08} 

In this paper, we present an extension to the nonequilibrium case of Monte Carlo approaches  based on an unbiased sampling of diagrammatic expansions\cite{Rubtsov05, Werner06, Werner06Kondo, Gull08} which, for equilibrium properties, have been shown to be powerful enough to access extremely low temperatures and flexible enough to treat a wide range of Hamiltonians. One of the specific implementations we present is  closely related to recent work by M\"uhlbacher and Rabani \cite{Muehlbacher08} and Schiro and Fabrizio \cite{Schiro08} for non-interacting dots with coupling to phonons. An extension to interacting dots has been employed by Schmidt {\it et al.} in Ref.~[\onlinecite{Schmidt08}]. 
Here, we provide a systematic analysis of the real-time diagrammatic approach, including a discussion of the strengths and weaknesses of these methods, and the regimes in which accurate results can be obtained. 
We discuss the formalism and details of the measurement formulae and implementations, and present results for observables including dot double occupancy and current through the dot. The rest of this paper is organized as follows: in Section \ref{Formalism} we outline the formalism we use and specify the model we treat, in Section {\ref{weak}} we present the weak-coupling formalism and in Section {\ref{strong}} the ``strong coupling" or hybridization expansion method. Section {\ref{results}} presents results for the time dependent dot occupation and double occupancy, 
and Section {\ref{current} discusses the  current.
Section {\ref{conclusions}} is a conclusion and outlook. An Appendix presents derivations of some needed formulae.

\section{Formalism and Model \label{Formalism}}

\subsection{General considerations}

A quantum impurity model is described by  the Hamiltonian
\begin{equation}
H_{QI}=H_\text{dot}+H_\text{bath}+H_\text{mix}.
\label{HQI}
\end{equation}
Here $H_\text{dot}$ describes a system with a finite dimensional Hilbert space, which we refer to as the ``impurity'', or ``dot'', $H_\text{bath}$ describes one or more infinite reservoirs characterized by a continuum of levels, and $H_\text{mix}$ the coupling between the impurity and the reservoirs. We assume that at time $t=0$ the state of the system is given by a density matrix $\rho_0$ which will be specified in detail later.  The statement that $H_\text{bath}$ is an infinite reservoir implies that the distribution function describing the occupation of the energy levels of $H_\text{bath}$ is independent of the coupling to  the  dot.  
\begin{figure}[t]
\begin{center}
\includegraphics[angle=0, width=0.5\columnwidth]{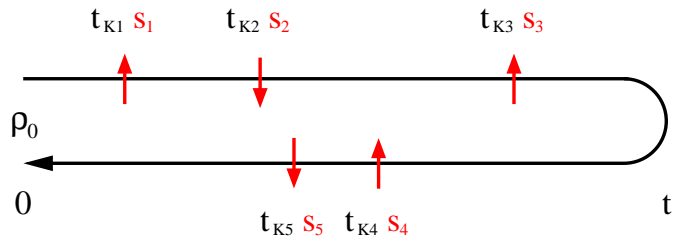}
\caption{
Example of a Monte Carlo configuration corresponding to perturbation order $n=5$ and $n_+=3$, $n_-=2$. 
}
\label{contour}
\end{center}
\end{figure}

The theoretical task is to evaluate the expectation value $\left<{\cal O}(t)\right>$ of an operator ${\cal O}$ at time $t$, i.e. to compute
\begin{equation}
\langle {\cal O} (t) \rangle =Tr\Big[\rho_0 e^{i\int_0^t dt' H_{QI}(t')}{\cal O} e^{-i\int_0^t dt'' H_{QI}(t'')}\Big]
\label{Odef}
\end{equation}
(the generalization to operators with multiple time dependences is straightforward and will not be written explicitly).  For a system in thermal equilibrium the issues in computing $\left<{\cal O}\right>$ are well understood. In this paper we are concerned with numerical approaches to describing the nonequilibrium situation. Nonequilibrium may enter through a time dependence of parameters in $H_{QI}$ (``irradiation"), through the correlators of the operators in $H_\text{bath}$ (``nonequilibrium reservoirs'') or through  an initial density matrix $\rho_0$ which is different from the long-time limit.  Our explicit considerations in this paper pertain mainly to the ``nonequilibrium reservoirs'' 
and ``nonequilibrium $\rho_0$''
case, but the methods we present generalize straightforwardly to other situations.

One may \cite{Kadanoff} view the expectation value in Eq.~(\ref{Odef}) as an evolution on the Schwinger-Keldysh contour illustrated in Fig.~\ref{contour}  from time $t=0$ (when the system is described by the density matrix $\rho_0$) to time $t$ (at which the operator is measured), and then back to time $0$. Our general strategy for evaluating Eq.~(\ref{Odef}) is to write $H_{QI}$ as a sum of two terms: one, $H_0$ for which the time evolution can be treated exactly and another, $H_I$, which is treated by a formal perturbative expansion. The expansion in $H_I$ generates a series of diagrams which are sampled stochastically, using an importance sampling which accepts or rejects proposed diagrams on the basis of their contributions to $\langle \tilde{ \cal O}\rangle$ with, for example, $\tilde {\cal O}=1$.  Two forms of expansion are considered: One is a ``weak coupling'' method, in which $H_\text{dot}$ is partitioned into a quadratic part $H_\text{dot}^0$ and an interacting part $H_U$, the combination $H_\text{dot}^0+H_\text{mix}+H_\text{bath}$ is diagonalized, $\rho_0$ is taken to be the corresponding density matrix, and the expansion is constructed in terms of $H_U$. The other is a ``strong coupling'' (more properly, ``hybridization'') expansion in which $H_\text{dot}$ and $H_\text{bath}$ are treated exactly, $\rho_0$  is the density matrix corresponding to the direct product of a lead density matrix and a density matrix describing the dot decoupled from the leads, and $H_\text{mix}$ is treated as a perturbation.  The hybridization expansion for nonequilibrium problems was previously presented by M\"uhlbacher and Rabani \cite{Muehlbacher08} in the context of noninteracting electrons coupled to phonons, and has been applied to interacting dots in Ref.~[\onlinecite{Schmidt08}]. An essentially identical formalism has also been 
discussed in Ref.~[\onlinecite{Schiro08}]. 

Methods based on stochastically sampled diagrammatic expansions have had considerable success in equilibrium quantum impurity problems at temperature $T>0$.\cite{Rubtsov05, Werner06, Werner06Kondo, Gull08}  There, the expansion can be formulated on the imaginary time axis $0\le \tau < 1/T$ (only one contour is needed) and the expansion parameter is $-H_I(\tau)=e^{\tau H_0}(-H_I)e^{-\tau H_0}$. The fermionic sign problem can be avoided (at least for sufficiently small dots) and temperatures as low as $0.1\%$ of the basic scales in the problem can be reached without inordinate effort.
Three related sources of difficulty arise in the nonequilibrium problem. First, the expansion must be done for real times, so convergence of the perturbation theory is oscillatory rather than exponential: diagrams come with factors of $i$ to powers relating to the perturbation order. Second, two contours rather than one are required, doubling the perturbation order required to reach a given time. Third, in nonequilibrium situations the form of the density matrix is crucial to the quantities (such as the current) which are to be computed; thus it is essential that the computation proceed for long enough to build up the correct entanglement between the impurity and the bath. All of these factors limit the range over which accurate results can be obtained, but the crucial constraint is the dynamical sign problem resulting from the oscillatory convergence.

\subsection{Model}

The results in this paper are presented for the simplest possible situation, a ``dot" consisting of a single spin-degenerate ($\sigma$) level with a Hubbard interaction $U$,  coupled by hybridization $V$  to two reservoirs (``leads") labeled by $\alpha=L,R$, with nonequilibrium entering via a possible difference between reservoir chemical potentials. The extension to more general situations is straightforward
and involves no new conceptual issues.

The Hamiltonian we consider is 
\begin{eqnarray}
H_\text{bath} &=& \sum_{\alpha=L,R} \sum_{p,\sigma} \big(\epsilon^\alpha_{p,\sigma}-\mu_\alpha \big)a^{\alpha \dagger}_{p,\sigma} a^\alpha_{p,\sigma},\label{H_bath}\\
H_\text{mix} &=& \sum_{\alpha=L,R} \sum_{p,\sigma} \big(V_p^\alpha a^{\alpha \dagger}_{p,\sigma}d_\sigma+h.c. \big),\label{H_mix}\\
H^0_\text{dot}&=&(\epsilon_d+U/2)\sum_\sigma n_{d,\sigma},\label{H_d}\\
H_{U} &=& U(n_{d,\uparrow} n_{d,\downarrow}-(n_{d,\uparrow}+n_{d,\downarrow})/2).\label{H_U}
\end{eqnarray}
It is also convenient to define
\begin{equation}
H_\text{dot} = H^0_\text{dot}+H_U.
\label{HDOT}
\end{equation}
The initial density matrix  is such that the correlators of lead operators are ($f_T(x)=(e^{x/T}+1)^{-1}$ is the Fermi distribution function for temperature $T$)
\begin{equation}
\left<a^{\alpha\dagger}_{p,\sigma}a^{\beta}_{p',\sigma'}\right>=\delta_{\alpha,\beta}\delta_{p,p'}\delta_{\sigma,\sigma^{'}}f_{T_\alpha}(\epsilon^\alpha_{p,\sigma}-\mu_\alpha)
\label{bathdistribution}
\end{equation}
and the statement that $H_\text{bath}$ describes infinite reservoirs is the statement that Eq.~(\ref{bathdistribution}) holds at all times. 

The model has three important energy scales: $\epsilon_d$ which controls the steady state dot occupancy, the interaction scale $U$, and the level broadening 
\begin{equation}
\Gamma^\alpha(\omega)=\pi\sum_p|V_p^\alpha|^2\delta(\omega-\epsilon_p^\alpha)
\label{Gamdef}
\end{equation}
associated with lead $\alpha$. The total level broadening is 
\begin{equation}
\Gamma=\Gamma^L+\Gamma^R
\label{Gammatotal}
\end{equation}
and the dimensionless measure of interaction strength is $U/\Gamma$. Very roughly, strong coupling physics appears for $U\gtrsim \pi \Gamma$ while the opposite limit is reasonably well described by perturbation theory in $U$ (see Section \ref{results}).

\section{Weak-coupling algorithm \label{weak}}

\subsection{Weak-coupling expansion and auxiliary field decomposition}

In the weak coupling expansion we treat $H_0\equiv H^0_\text{dot}+H_\text{mix}+H_\text{bath}$ exactly and $H_U$ as a perturbation. $H_0$ is a noninteracting problem for which the density matrix and all correlators of the dot-lead system can be determined exactly.  We take the initial density matrix to be the steady-state  density matrix corresponding to $H_0$ (here we assume the temperatures of the two leads are identical; the generalization to unequal temperatures is straightforward)
\begin{equation}
\rho_0=\frac{e^{-\beta H_0}}{Tr e^{-\beta H_0}},
\label{rho0weakcoupling}
\end{equation}
and consider  the interaction to be  turned on at time $t=0$.

We formulate the perturbation theory in $U$ as a real-time incarnation of the recently developed continuous-time auxiliary field method of Ref.~[\onlinecite{Gull08}], which itself is an adaptation of ideas in Ref.~[\onlinecite{Rombouts99}] to impurity models. 
The starting point for the real-time auxiliary field method is the following expression for the identity:
\begin{equation}
1 = Tr \rho_0 e^{it(H_0+H_U-K/t)} e^{-it(H_0+H_U-K/t)},
\label{identity}
\end{equation}
with $K$ 
a constant which is in principle arbitrary and may be chosen to optimize the simulation. As discussed below we find that choosing $K$ to be negative, and small in magnitude appears to work best.
Using an interaction representation in which the time evolution of the operators is given by $O(s)=e^{isH_0}Oe^{-isH_0}$ we can rewrite Eq.~(\ref{identity}) as
\begin{equation}
1=Tr \rho_0 \Big(\tilde T e^{i\int_0^t ds (H_U(s)-K/t)}\Big)e^{itH_0} e^{-itH_0} \Big(T e^{-i\int_0^t ds (H_U(s)-K/t)}\Big),
\label{identity2}
\end{equation}
with $T$ the time ordering and $\tilde T$ the anti-time ordering operator, and expand the time ordered exponentials into a power series. This leads to the expression
\begin{eqnarray}
1 &=&Tr \rho_0
\sum_{m} (-iK/t)^m \int_0^t d\tilde t_1 \ldots \int_{\tilde t_{m-1}}^t d\tilde t_m e^{i \tilde t_1 H_0}(1-tH_U/K)\ldots e^{i(\tilde t_{m}-\tilde t_{m-1})H_0}(1-tH_U/K)e^{i(t-\tilde t_m)H_0}\nonumber\\
&\times& \sum_{n} (iK/t)^n \int_0^t d t_1 \ldots \int_{ t_{n-1}}^t d t_n e^{-i(t- t_n)H_0}(1-tH_U/K)\ldots e^{-i(t_2-t_1)H_0}(1-tH_U/K)e^{-it_1H_0}.
\end{eqnarray}
Using the explicit form for $H_U$ (Eq.~(\ref{H_U})) and the auxiliary field decomposition of Ref.~[\onlinecite{Rombouts99}] we can rewrite the interaction term as 
\begin{eqnarray}
1-(t U/K)(n_{d,\uparrow} n_{d,\downarrow}-(n_{d,\uparrow}+n_{d,\downarrow})/2)) &=& 1/2\sum_{s=-1,1}e^{\gamma s (n_{d,\uparrow}-n_{d,\downarrow})},\\
\cosh(\gamma)&=&1+(tU)/(2K).
\label{decouple}
\end{eqnarray}
Note that the constant $K$ has been introduced to enable this decomposition. 
The trace is now a product of exponentials of one-body operators, 
\begin{eqnarray}
1&=& \sum_m \sum_n (-i)^m i^n (K/2t)^{m+n} \sum_{\tilde s_1,\ldots,\tilde s_n} \sum_{s_1,\ldots,s_m}\int_0^t d\tilde t_{1} \ldots \int_{\tilde t_{m-1}}^t d\tilde t_m \int_0^t dt_1 \ldots \int_{t_{n-1}}^t dt_n \prod_\sigma (1/Tr e^{-\beta H_{0,\sigma}})\nonumber\\
&\times& Tr \Big[ e^{-\beta H_{0,\sigma}}
e^{i\tilde t_1H_{0,\sigma}} e^{\gamma \tilde s_1 \sigma n_{d,\sigma}} \ldots e^{i(\tilde t_{m}-\tilde t_{m-1})H_{0,\sigma}} e^{\gamma \tilde s_m \sigma n_{d,\sigma}}e^{-i(\tilde t_m-t_n)H_{0,\sigma}} 
e^{\gamma s_n \sigma n_{d,\sigma}} \ldots e^{-i(t_2-t_1)H_{0,\sigma}} e^{\gamma s_1 \sigma n_{d,\sigma}}e^{-i t_1H_{0,\sigma}} \Big],\nonumber
\end{eqnarray}
and can be expressed \cite{Gull08} in terms of  determinants of two $(n+m)\times(n+m)$ matrices 
\begin{equation}
N_\sigma^{-1} = e^{S_\sigma}-iG_{0,\sigma}(e^{S_\sigma}-I) 
\end{equation}
as
\begin{eqnarray}
1&=& \sum_m \sum_n (-i)^m i^n (K/2t)^{m+n} \sum_{\tilde s_1,\ldots,\tilde s_n} \sum_{s_1,\ldots,s_m} \int_0^t d\tilde t_{1} \ldots \int_{\tilde t_{m-1}}^t d\tilde t_m \int_0^t dt_1 \ldots \int_{t_{n-1}}^t dt_n \prod_\sigma \det N_\sigma^{-1},
\end{eqnarray}
with $e^{S_\sigma}=\text{diag}(e^{\gamma \tilde s_1\sigma}, \ldots, e^{\gamma \tilde s_m \sigma}, e^{\gamma s_n\sigma}, \ldots, e^{\gamma s_1 \sigma})$, and $G_{0,\sigma}$ given by
\begin{equation}
G_{0,\sigma}(t_K',t_K'')=\left\{
\begin{array}{ll}
G_{0,\sigma}^<(t',t''), & t'_K < t''_K\\
G_{0,\sigma}^>(t',t''), & t'_K \ge t''_K\label{eqn:G0input}
\end{array}
\right. .
\end{equation}
In the above expression, $G^<_0(t,t')=i\langle d^\dagger(t')d(t)\rangle_0$, $G^>_0(t,t')=-i\langle d(t)d^\dagger(t')\rangle_0$, $t_K$ is the ``Keldysh time" coordinate along the unfolded Keldysh contour (Fig.~\ref{contour}) and $t$ the time corresponding to $t_K$.  These Green's functions may be computed by standard methods.\cite{Rammer96} A general expression is presented in Appendix A; for our actual computations we will use the infinite bandwidth limit in which the level broadening is independent of $\omega$ so that 
\begin{eqnarray}
G_0^{</>}(t',t'')&=&\pm i\sum_{\alpha=L,R}\Gamma^\alpha\int \frac{d\omega}{2\pi}e^{-i\omega(t'-t'')}\frac{1\mp  \tanh\left(\frac{\omega-\mu_\alpha}{2T}\right)}{(\omega-\epsilon_d-U/2)^2+\Gamma^2}
\end{eqnarray}
with the upper sign pertaining to $G_0^<$ and the lower sign to $G_0^>$.

\subsection{Detailed balance and fast updates}

The algorithm samples  auxiliary Ising spin configurations $\{ (t_{K,1}, s_1),(t_{K,2}, s_2), \ldots (t_{K,n}, s_n) \}$ time ordered along  the ``Keldysh" contour $0\rightarrow t \rightarrow 0$ (see Fig.~\ref{contour}) by random insertions and removals of spins.  The complex ``weight" of a spin configuration is given by
\begin{equation}
w(\{ (t_{K,1}, s_1),(t_{K,2}, s_2), \ldots (t_{K,n}, s_n) \})=(-i^{n_-})(i^{n_+})(K dt/2t)^{n_-+n_+}\prod_\sigma \det N_\sigma^{-1},
\label{weight}
\end{equation}
where $n_+$ denotes the number of spins on the forward contour and $n_-$ the number of spins on the backward contour ($n=n_++n_-$).

The detailed balance condition for insertion/removal of a spin is similar to the imaginary time formulation of Ref.~[\onlinecite{Gull08}].  Assuming that we pick a random time on the unfolded contour of length $2t$ and a random direction for this new spin ($p^\text{prop}(n-1\rightarrow n)=(1/2)(dt/(2t)$), and propose to remove this spin with probability $p^\text{prop}(n\rightarrow n-1)=1/n$ we get
\begin{equation}
\frac{p(n-1\rightarrow n)}{p(n\rightarrow n-1)}=\pm i\frac{2K}{n}\prod_\sigma \frac{\det (N^{-1}_{n})_\sigma}{\det (N^{-1}_{n-1})_\sigma},
\end{equation}
with the factor $+i$ corresponding to a spin which is inserted on the forward contour and $-i$ to a spin which is inserted on the backward contour. 

For the fast updates, let us consider the most complicated case, which is the insertion of a spin. This update adds one row and one column to the $(n-1)\times(n-1)$ matrix $N$, resulting in the $n\times n$ matrix $N'$ (we assume here that this new row/column is the last one, $n$, and drop the spin index). The determinant ratio is
\begin{equation}
r=\frac{\det (N'^{-1})}{\det (N^{-1})} = (e^S-iG_0(e^S-I))_{n,n}-\sum_{i=1}^{n-1}R_i (e^S-iG_0(e^S-I))_{i,n},
\end{equation}
with $R_i=\sum_{j=1}^{n-1}(e^S-iG_0(e^S-I))_{n,j}N_{j,i}$. The calculation of this quantity requires $O(n^2)$ operations. The new matrix elements are given by
\begin{align}
N'_{i,j} &= N_{i,j}+\frac{1}{r}L_iR_j,\\
N'_{i,n} &= -\frac{1}{r} L_i,\\
N'_{n,j} &= -\frac{1}{r} R_j,\\
N'_{n,n} &= \frac{1}{r},\label{det_rat_rem}
\end{align}
with $i=1,\ldots, n-1$ and $L_i=\sum_{j=1}^{n-1}N_{i,j}(e^S-iG_0(e^S-I))_{j,n}$. 

From Eq.~(\ref{det_rat_rem}) it follows that computing the determinant ratio for removing a spin is $O(1)$. The elements of the reduced matrix are obtained as
\begin{equation}
N_{i,j} = N'_{i,j}-\frac{N'_{i,n}N'_{n,j}}{N'_{n,n}}.
\end{equation}

\subsection{\bf Green's function, dot population and double occupancy}

To measure the Green's function $G_\sigma(t_K', t_K'')$ we have to insert an operator $d_\sigma$ at time $t_K'$ and an operator $d_\sigma^\dagger$ at time $t_K''$. The weights of these configurations $w(\{ (t_{K,1}, s_1), \ldots (t_{K,n}, s_n) \}; d_\sigma(t_K')d_\sigma^\dagger(t_K''))$ are related to those defined in Eq.~(\ref{weight}) by
\begin{equation}
\frac{w(\{ (t_{K,1}, s_1), \ldots (t_{K,n}, s_n) \}; d(t')d^\dagger(t''))}{w(\{ (t_{K,1}, s_1), \ldots (t_{K,n}, s_n) \})}=
\frac{1}{\det N_\sigma^{-1}}
\det \left(
\begin{array}{l|l}
N_\sigma^{-1}(i,j)&iG_{0,\sigma}(t_{K,i},t_K'')\\
\hline
-i G_{0,\sigma}(t_K',t_{K,j})(e^{\gamma\sigma s_j}-1)&iG_{0,\sigma}(t_K',t_K'')
\end{array}
\right).
\end{equation}
Hence, the Green's function can be obtained as the Monte Carlo average of the quantity (see also Ref.~[\onlinecite{Gull08}])
\begin{equation}
\tilde G_\sigma(t_K',t_K'')=G_{0,\sigma}(t_K',t_K'')+i\sum_{i,j=1}^n G_{0,\sigma}(t_K',t_{K,i})[(e^{S_\sigma}-1)N_\sigma]_{i,j}G_{0,\sigma}(t_{K,j}, t_K''),
\end{equation}
which yields the measurement formulas
\begin{eqnarray}
G_\sigma(t_K',t_K'') &=& \langle  \tilde G_\sigma(t_K',t_K'') \rangle,\\
n_\sigma(t_K) &=& 1-i\langle \tilde G_\sigma(t_K,t_K) \rangle,\\
n_\uparrow n_\downarrow (t_K) &=& \langle(1-i \tilde G_\uparrow(t_K,t_K)) (1-i \tilde G_\downarrow(t_K,t_K))\rangle.
\end{eqnarray}

\subsection{Current measurement}

The current from the dot to the left lead is 

\begin{equation}
I_L=\sum_\sigma I_{L\sigma}=-2\text{Im} \sum_\sigma \sum_{p\in L} V^L_{p,\sigma} \langle a^{L\dagger}_{p,\sigma}d_\sigma \rangle.
\end{equation}
Thus, in terms of the composite lead operator $\tilde a^\dagger_{L,\sigma}\equiv\sum_{p\in L} V^L_{p,\sigma} a^{L\dagger}_{p,\sigma}$, we find
\begin{align}
I_{L\sigma}(t)&=-2\text{Im} Tr \rho_0 \Big(\tilde T e^{i\int_0^t ds (H_I(s)-K/t)}\Big)e^{itH_0} \tilde a^\dagger_{L,\sigma} d_\sigma e^{-itH_0} \Big(T e^{-i\int_0^t ds (H_I(s)-K/t)}\Big)\nonumber\\
&= -2\text{Im}\sum_m \sum_n (-i)^m i^n (K/2t)^{m+n} \sum_{\tilde s_1,\ldots,\tilde s_n} \sum_{s_1,\ldots,s_m}\int_0^t d\tilde t_{1} \ldots \int_{\tilde t_{m-1}}^t d\tilde t_m \int_0^t d t_1 \ldots \int_{t_{n-1}}^t dt_n \det N_{\bar\sigma}^{-1} \frac{1}{Tr e^{-\beta H_{0,\sigma}}}\nonumber\\
&\times Tr \Big[ e^{-\beta H_0}
e^{i\tilde t_1H_{0,\sigma}} e^{\gamma \tilde s_1 \sigma n_{d,\sigma}} \ldots e^{\gamma \tilde s_m \sigma n_{d,\sigma}}
e^{i(t-\tilde t_m)H_{0,\sigma}}
\tilde a^\dagger_{L,\sigma} d_\sigma
e^{-i(t-t_n)H_{0,\sigma}}
e^{\gamma \sigma s_n \sigma n_{d,\sigma}} \ldots 
e^{\gamma s_1 \sigma n_{d,\sigma}}e^{-i t_1H_{0,\sigma}} \Big],\nonumber\\
\label{current_long}
\end{align}
with $\bar \sigma$ the spin which is opposite to $\sigma$ (this spin component has no operator $\tilde a$ and thus simply gives the usual factor $\det N_{\bar \sigma}^{-1}$). The measurement of the current is thus very similar to the measurement of the Green's functions, but one factor in the Wick decomposition is now %
\begin{equation}
A(t_K,t'_K) = \left\{ 
\begin{array}{ll}
A^<(t,t')\equiv \langle \tilde a^{L\dagger}_{\sigma}(t') d_\sigma(t) \rangle_0, &t_K\le t'_K\\
A^>(t,t')\equiv -\langle d_\sigma(t)  \tilde a^{L\dagger}_{\sigma}(t') \rangle_0, &t_K>t'_K\\
\end{array}
\right. ,
\end{equation}
A derivation and a general expression are given in Appendix A. In the infinite bandwidth limit we have 
\begin{eqnarray}
\left.
\begin{array}{ll}
A^<(t,t')\\
A^>(t,t')\\
\end{array}
\right\}
&=&-2i\int \frac{d\omega}{2\pi}e^{-i\omega(t-t')}\frac{\Gamma_L \Gamma_R (f(\omega-\mu_L)-f(\omega-\mu_R))}{(\omega - \epsilon_d-U/2)^2+\Gamma^2} \nonumber\\
&&+2\Gamma_L\int \frac{d\omega}{2\pi}e^{-i\omega(t-t')}\frac{(\omega - \epsilon_d-U/2)}{(\omega - \epsilon_d-U/2)^2+\Gamma^2} 
\times \left\{
\begin{array}{ll}
f(\omega-\mu_L)\\
(f(\omega-\mu_L)-1)\\
\end{array}
\right. .
\label{A}
\end{eqnarray}

The trace factor in Eq.~(\ref{current_long}) for an $n$-th order diagram corresponding to the $n\times n$ matrix $N_\sigma^{-1}$ is the determinant of the $(n+1)\times(n+1)$ matrix 
\begin{equation}
M^{-1}_\sigma = 
\left(
\begin{array}{l|l}
N_\sigma^{-1}(i,j)&A(t_{K,i},t)\\
\hline
-i G_{0,\sigma}(t,t_{K,j})(e^{\gamma\sigma s_j}-1)&A(t,t)
\end{array}
\right).
\end{equation}
The current can thus be expressed as follows:
\begin{equation}
I_L=-2\text{Im} \sum_\sigma\sum_c w^{I_\sigma}_{c}=-2\text{Im}\sum_\sigma\Bigg[\frac{\sum_c |w_c| (w^{I_\sigma}_{c}/|w_c|)}{\sum_c |w_c| \phi_c}\Bigg]=-2\text{Im}\sum_\sigma\Bigg[\Big\langle \frac{w_{c}^{I_\sigma}}{|w_c|}\Big\rangle_{|w_c|}\frac{1}{\langle\phi_c\rangle_{|w_c|}}\Bigg],\label{I_weight}
\end{equation}
with $\phi_c$ the phase of the weight $w_c$ (Eq.~(\ref{weight})) and
\begin{eqnarray}
\frac{w^{I_\sigma}_{c}}{w_c}&=&\frac{\det N_{\bar\sigma}^{-1}\det M_\sigma^{-1}}{\det N_{\bar \sigma}^{-1}\det N^{-1}_\sigma}=A(t,t)+\sum_{n,m} i G_{0,\sigma}(t,t_{K,n}) 
[(e^{S_\sigma}-1)N_\sigma]_{n,m}
A(t_{K,m},t).\label{weight_rat}
\end{eqnarray}
Combining Eqs. (\ref{I_weight}) and (\ref{weight_rat}), the current measurement formula becomes
\begin{align}
I=I_L =&-2\text{Im}\sum_\sigma \Big[ A(t,t)+\Big\langle \sum_{n,m} i G_{0,\sigma}(t,t_{K,n}) [(e^{S_\sigma}-1)N_\sigma]_{n,m}A(t_{K,m},t) \phi_c\Big\rangle_{|w_c|}\frac{1}{\langle\phi_c\rangle_{|w_c|}}\Big].
\end{align}
The first term in this expression is the steady-state current for the non-interacting system
\begin{equation}
I_0 = -2\text{Im} (2 A(t,t)) = 8 \int \frac{d\omega}{2\pi}\frac{\Gamma_L \Gamma_R (f(\omega-\mu_L)-f(\omega-\mu_R))}{(\omega - \epsilon_d-U/2)^2+\Gamma^2}.
\label{I_0}
\end{equation}

\subsection{Real-time Hirsch-Fye method}

In order to assess the efficiency of the continuous-time weak-coupling approach, we have also performed calculations using the real-time version of the Hirsch-Fye method.\cite{HirschFye86} In this method, time is discretized along the Schwinger-Keldysh contour and the identity is expressed as 
\begin{eqnarray}
1&=&\mbox{Tr}\rho_0e^{itH}e^{-itH}=
\mbox{Tr}\rho_0\prod_{l=1}^{L/2}e^{i\Delta t[H_0+H_U]}
\prod_{l=1}^{L/2}e^{-i\Delta t[H_0+H_U]}\\
&\simeq&
\mbox{Tr}\rho_0\prod_{l=1}^{L/2}e^{i\Delta t H_0}
e^{i\Delta t H_U}
\prod_{l=1}^{L/2}e^{-i\Delta t H_0}e^{-i\Delta t H_U},
\end{eqnarray}
where we used the Trotter breakup and $L$ denotes the (even) number of time slices. In the real time Hirsch-Fye method, the interaction term for each time slice is decoupled using the Hubbard-Stratonovich transformation 
\begin{equation}
e^{-i\Delta tH_U}=\frac{1}{2}\sum_{s=\pm 1}
e^{\lambda s(n_\uparrow-n_\downarrow)},\qquad
\lambda=\cosh^{-1}(e^{i\Delta tU/2}).
\end{equation}
After this decoupling, we obtain a product of exponentials of one-body operators and the trace can thus be computed analytically. Besides the fact that $\lambda$ is a complex number and that the non-interacting Green's function is given by Eq.~(\ref{eqn:G0input}), the derivation of the algorithm and the sampling procedure are identical to the original imaginary-time Hirsch-Fye method. Two sources of error exist in this method. One is the discretization error due to the Trotter breakup and the other is the stochastic error which becomes severe at large $L$ due to the sign problem. Because of these limitations, the real-time Hirsch-Fye method is restricted to short time calculations, 
and indeed the time limits appear to be more stringent than in the continuous-time methods we introduce here
(see 
for example Fig. \ref{double_weak}).

\section{Hybridization-expansion Algorithm \label{strong}}
\subsection{Formalism}

A complementary diagrammatic Monte Carlo algorithm can be obtained by performing an expansion in powers of the dot-lead hybridizations $V$. This simulation approach has been introduced for equilibrium systems (imaginary-time formalism) in Refs.~[\onlinecite{Werner06, Werner06Kondo, Werner07Holstein}] and was recently discussed for a nonequilibrium dot with phonons (but without electron-electron interactions) in Refs.~[\onlinecite{Muehlbacher08, Schiro08}]. It has been applied to interacting dots in Ref.~[\onlinecite{Schmidt08}]. We will present here the derivation for the impurity model defined in Eqs.~(\ref{H_bath})-(\ref{H_U}), but the method can easily be extended to general classes of impurity models by using the matrix formulation of Ref.~[\onlinecite{Werner06Kondo}].

In the hybridization expansion approach one adopts an interaction representation with respect to the dot-lead mixing, so the time evolution of the operators is given  by the local part of the Hamiltonian, $H_\text{loc}=H_\text{dot}+H_\text{bath}$, and the starting point is the identity
\begin{equation}
1=Tr \rho_0 \Big(\tilde T e^{i\int_0^t ds H_\text{mix}(s)}\Big)e^{it H_\text{loc}}e^{-it H_\text{loc}}\Big(T e^{-i\int_0^t ds H_\text{mix}(s)}\Big).
\end{equation}
The initial state of the system is specified by the density matrix $\rho_0=\rho_\text{dot}\otimes\rho_\text{bath}$, with $\rho_\text{bath}$ a function of inverse temperature $\beta$ and the chemical potentials $\mu_{L,R}$. In the calculations presented here we assume that the dot is initially empty, $\rho_\text{imp}=|0\rangle\langle 0|$. 

Expanding the time ordered exponentials into a power series yields
\begin{eqnarray}
1&=&Tr \rho_0 \sum_m i^m \int_0^t d\tilde t_1\ldots \int_{\tilde t_{m-1}}^t d\tilde t_m H_\text{mix}(\tilde t_1)\ldots H_\text{mix}(\tilde t_m)\nonumber\\
&\times& \sum_n (-i)^n \int_0^t dt_1\ldots \int_{t_{n-1}}^t dt_n H_\text{mix}(t_n)\ldots H_\text{mix}(t_1).
\end{eqnarray}
Because $H_\text{mix}=\sum_\sigma (H_\text{mix}^{d_\sigma}+H_\text{mix}^{d^\dagger_\sigma})$ with $H_\text{mix}^{d_\sigma}=\sum_{\alpha=L,R} \sum_{p} V_p^\alpha a^{\alpha \dagger}_{p,\sigma}d_\sigma$, $H_\text{mix}^{d^\dagger_\sigma}=(H_\text{mix}^{d_\sigma})^\dagger$ and the time evolution conserves the spin, we need for each $\sigma$ separately an equal number of creation and annihilation operators on the Keldysh contour $0\rightarrow t\rightarrow 0$:
\begin{eqnarray}
1&=&\sum_{m_\sigma+n_\sigma=m_\sigma'+n_\sigma'} \prod_\sigma i^{m_\sigma+m_\sigma'}(-i)^{n_\sigma+n_\sigma'} \nonumber\\
&\times&\int_0^t d\tilde t^\sigma_1\ldots \int_{\tilde t^\sigma_{m_\sigma-1}}^t d\tilde t^\sigma_{m_\sigma}
\int_0^t d\tilde t'^\sigma_1\ldots \int_{\tilde t'^\sigma_{m'_\sigma-1}}^t d\tilde t'^\sigma_{m'_\sigma}
\int_0^t dt^\sigma_1\ldots \int_{t^\sigma_{n_\sigma-1}}^t dt^\sigma_{n_\sigma}
\int_0^t dt'^\sigma_1\ldots \int_{t'^\sigma_{n'_\sigma-1}}^t dt'^\sigma_{n'_\sigma}\nonumber\\
&\times& Tr \Bigg[\rho_0 \tilde T T \prod_\sigma H^{d_\sigma}_\text{mix}(\tilde t^\sigma_1)H^{d_\sigma^\dagger}_\text{mix}(\tilde t'^\sigma_1) H^{d_\sigma}_\text{mix}(\tilde t^\sigma_2)H^{d_\sigma^\dagger}_\text{mix}(\tilde t'^\sigma_2) \ldots e^{iH_\text{loc}t}e^{-iH_\text{loc}t}\ldots
H^{d_\sigma}_\text{mix}(t^\sigma_2)H^{d_\sigma^\dagger}_\text{mix}(t'^\sigma_2) H^{d_\sigma}_\text{mix}(t^\sigma_1)H^{d_\sigma^\dagger}_\text{mix}(t'^\sigma_1)\Bigg],\nonumber\\
\end{eqnarray}
where $\tilde T$ is the anti-time ordering operator for the $\tilde t$s and $T$ the time ordering operator for the $t$s. At this stage we can separate the bath operators $a_{p,\sigma}^\alpha$ from the dot operators $d_\sigma$ and write
\begin{eqnarray}
1&=&\sum_{m_\sigma+n_\sigma=m_\sigma'+n_\sigma'} \prod_\sigma i^{m_\sigma+m_\sigma'}(-i)^{n_\sigma+n_\sigma'} \nonumber\\
&\times&\int_0^t d\tilde t^\sigma_1\ldots \int_{\tilde t^\sigma_{m_\sigma-1}}^t d\tilde t^\sigma_{m_\sigma}
\int_0^t d\tilde t'^\sigma_1\ldots \int_{\tilde t'^\sigma_{m'_\sigma-1}}^t d\tilde t'^\sigma_{m'_\sigma}
\int_0^t dt^\sigma_1\ldots \int_{t^\sigma_{n_\sigma-1}}^t dt^\sigma_{n_\sigma}
\int_0^t dt'^\sigma_1\ldots \int_{t'^\sigma_{n'_\sigma-1}}^t dt'^\sigma_{n'_\sigma}\nonumber\\
&\times& Tr_d \Bigg[\rho_\text{dot} \tilde T T \prod_\sigma d_\sigma(\tilde t^\sigma_1)d_\sigma^\dagger(\tilde t'^\sigma_1) d_\sigma(\tilde t^\sigma_2) d_\sigma^\dagger(\tilde t'^\sigma_2) \ldots e^{iH_\text{dot}t}e^{-iH_\text{dot}t}\ldots
d_\sigma(t^\sigma_2)d_\sigma^\dagger(t'^\sigma_2) d_\sigma(t^\sigma_1)d_\sigma^\dagger(t'^\sigma_1)\Bigg]\nonumber\\
&\times& Tr_\text{bath} \Bigg[\rho_\text{bath} \tilde T T \prod_\sigma 
\sum_{\tilde p_1\tilde \alpha_1;...;\tilde p_{m_\sigma}\tilde \alpha_{m_\sigma}}\sum_{\tilde p'_1\tilde \alpha'_1;...;\tilde p'_{m'_\sigma}\tilde \alpha'_{m'_\sigma}}
\sum_{p_1\alpha_1;...;p_{n_\sigma}\alpha_{n_\sigma}}\sum_{p'_1\alpha'_1;...;p'_{n'_\sigma}\alpha'_{n'_\sigma}}
V_{\tilde p_1}^{\tilde \alpha_1} V_{\tilde p'_1}^{\tilde \alpha'_1*}
\ldots V_{p_1}^{\alpha_1} V_{p'_1}^{\alpha'_1*}\nonumber\\
&&\hspace{12mm}a^\dagger_\sigma(\tilde t^\sigma_1)a_\sigma(\tilde t'^\sigma_1) a^\dagger_\sigma(\tilde t^\sigma_2) a_\sigma(\tilde t'^\sigma_2) \ldots e^{iH_\text{bath}t}e^{-iH_\text{bath}t}\ldots
a^\dagger_\sigma(t^\sigma_2)a_\sigma(t'^\sigma_2) a^\dagger_\sigma(t^\sigma_1)a_\sigma(t'^\sigma_1)\Bigg],
\label{expansion_complete}
\end{eqnarray}
with $\alpha_i\in\{L,R\}$.  Since the leads are non-interacting (see Eq.~(\ref{H_bath})) we can evaluate the factor $Tr_\text{bath}[\ldots]$ exactly. Due to Wick's theorem one obtains a product of two determinants $\prod_\sigma \det M^{-1}_\sigma$, with the size of $M_\sigma^{-1}$ given by the number of operators $d_\sigma$ on the Keldysh contour ($m_\sigma+n_\sigma$). The matrix elements are given by \cite{Muehlbacher08, Werner06Kondo}
\begin{equation}
M_\sigma^{-1}(i,j)=i\Delta(t^\sigma_{K,i},t'^\sigma_{K,j}),
\label{M_inv}
\end{equation}
where $t^\sigma_{K,i}$ denotes the position of the $i$th annihilation operator and $t'^\sigma_{K,j}$ the position of the $j$th creation operator for spin $\sigma$ on the unfolded Keldysh contour. The function $\Delta$ is discussed in Appendix $A$ and is 
\begin{equation}
\Delta(t_K,t_K')=\left\{
\begin{array}{ll}
\Delta^<(t'-t)\equiv\Delta_L^<(t'-t)+\Delta_R^<(t'-t) & t_K \ge t'_K,\\
\Delta^>(t'-t)\equiv\Delta_L^>(t'-t)+\Delta_R^>(t'-t) & t_K < t'_K\\
\end{array}
\right. ,
\end{equation}
with 
\begin{eqnarray}
\Delta^<_\alpha(t)&=&-2i\int_{-\infty}^\infty \frac{d\omega}{2\pi} e^{-i\omega t} \Gamma^\alpha(\omega)f(\omega-\mu_\alpha)\label{Sigma_lesser},\\
\Delta^>_\alpha(t)&=&2i\int_{-\infty}^\infty \frac{d\omega}{2\pi} e^{-i\omega t} \Gamma^\alpha(\omega)(1-f(\omega-\mu_\alpha))\label{Sigma_greater}.
\end{eqnarray}

\begin{figure}[t]
\begin{center}
\includegraphics[angle=0, width=0.6\columnwidth]{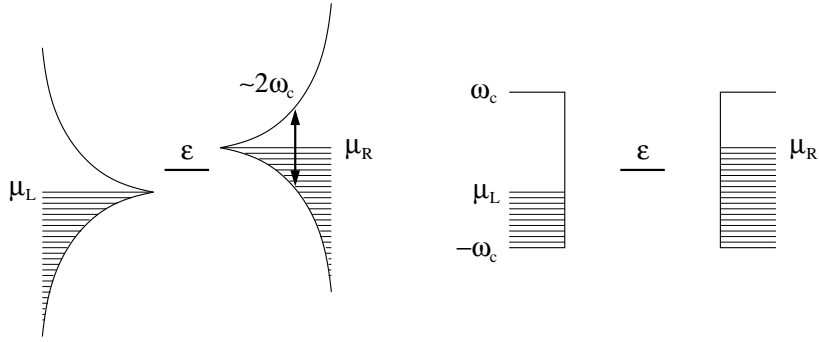}
\caption{Illustration of the band cutoffs considered in the simulations. The soft cutoff (left panel) is an exponentially decaying $\Gamma(\omega)$ which is centered at the chemical potential. The band with hard cutoff (right panel) is symmetric around $\omega=0$ and does not shift with the chemical potential.}
\label{dos}
\end{center}
\end{figure}

We give here simple expressions for two types of bands, illustrated in Fig.~\ref{dos},
which are exact in the limit $T\rightarrow 0$ and a good approximation for $T\ll \omega_c$.
The first is a band with soft cutoff, centered on the chemical potential,
\begin{equation}
\Gamma_\text{soft}^\alpha(\omega)=\Gamma^\alpha e^{-|\omega-\mu_\alpha|/\omega_c}.
\end{equation} 
For symmetric voltage bias ($\mu_L=-\mu_R=V/2$) and symmetric couplings ($\Gamma_L=\Gamma_R$) we obtain
\begin{eqnarray}
\Delta_\text{soft}^{< / >}(t)&\simeq&\Gamma \frac{\cos(\frac{V}{2}t)}{\beta\sinh(\frac{\pi }{\beta}\left(t\pm i/\omega_c\right))}.
\end{eqnarray}
The second example is a flat band centered at zero with a hard (Fermi-function like) cutoff at $\omega=\pm\omega_c$,
\begin{equation}
\Gamma_\text{hard}^\alpha(\omega)=\frac{\Gamma^\alpha}{(1+e^{\nu(\omega-\omega_c)})(1+e^{-\nu(\omega+\omega_c)})},
\end{equation}
which yields
\begin{eqnarray}
\Delta_\text{hard}^{< / >}(t)&\simeq&\Gamma\Bigg(\frac{\cos(\frac{V}{2}t)}{\beta\sinh(\frac{\pi }{\beta}t)}-\frac{e^{\pm i\omega_c t}}{\nu\sinh(\frac{\pi }{\nu}t)}\Bigg).
\end{eqnarray}

To evaluate the trace over the impurity states in Eq.~(\ref{expansion_complete}), $Tr_d[\dots]$, it is useful to employ the segment representation introduced for impurity models with density density interactions in Ref.~[\onlinecite{Werner06}]. The sequence of dot creation and annihilation operators uniquely determines the occupation of the dot at each time, and we can represent the time evolution using collections of segments for spin up and down electrons as shown in Fig.~\ref{segments}. Each segment depicts a time interval for which an electron with corresponding spin resides on the dot. The trace over the impurity states can then simply be expressed as
\begin{equation}
Tr_d\Big[\ldots\Big] = \rho_\text{imp}(c)\exp\Big[-i\epsilon_d\sum_\sigma(l^\sigma_\text{forward}-l^\sigma_\text{backward})-iU(l^\text{overlap}_\text{forward}-l^\text{overlap}_\text{backward})\Big].
\end{equation} 
Here, $\rho_\text{imp}(c)$ is the element of the impurity density matrix which is compatible with the operator sequence $c=\{t_{K,1}^\sigma, \ldots, t_{K,m_\sigma+n_\sigma}^\sigma; t'^\sigma_{K,1}, \ldots, t'^\sigma_{K,m'_\sigma+n'_\sigma}\}$ (assumed here to be $1$ for configurations which start and end with an empty dot and zero otherwise), $l^\sigma$ the length of the segments for spin $\sigma$ and $l^\text{overlap}$ the length of the overlap between spin up and down segments. 

\begin{figure}[t]
\begin{center}
\includegraphics[angle=0, width=0.5\columnwidth]{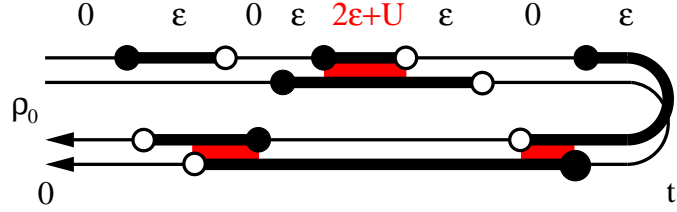}
\caption{Segment representation of a Monte Carlo configuration corresponding to perturbation order 4 for spin up (upper contour) and 2 for spin down (lower contour). Dot creation operators are shown as full circles and annihilation operators as open circles. The segments represent the time intervals in which an electron of the corresponding spin resides on the dot.}
\label{segments}
\end{center}
\end{figure}

\begin{figure}[t]
\begin{center}
\includegraphics[angle=0, width=0.5\columnwidth]{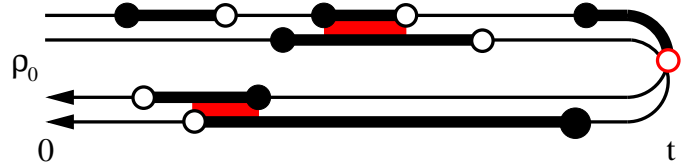}
\caption{Segment configurations obtained from the expansion of the current (Eq.~(\ref{current_hyb})) in powers of the dot-lead hybridization. There is a fixed operator $d_\sigma$ (red open circle) at time $t$ and the hybridization functions connecting to this operator have only a left ($L$) component.
}
\label{contour_current}
\end{center}
\end{figure}

Hence, the Monte Carlo simulation samples collections $c$ of segments on the doubled ``Keldysh contour" (one for each spin) according to their weight
\begin{eqnarray}
w(c) &=& \prod_\sigma i^{m_\sigma+m_\sigma'}(-i)^{n_\sigma+n_\sigma'} \det M_\sigma^{-1}dt^{m_\sigma+m_\sigma'+n_\sigma+n_\sigma'}\nonumber\\
&\times&\rho_\text{imp}(c)\exp\Big[-i\epsilon_d\sum_\sigma(l^\sigma_\text{forward}-l^\sigma_\text{backward})-iU(l^\text{overlap}_\text{forward}-l^\text{overlap}_\text{backward})\Big].
\label{weight_strong}
\end{eqnarray}
We implemented the following local updates of the segment configurations: i) insertion/removal of a segment, ii) insertion/removal of an anti-segment (empty space between segments) and iii) shifts of segment end-points. In order to use fast update formulas similar to those discussed in Section \ref{weak} we store and manipulate the matrices $M_\sigma$, that is, the inverse of the matrices defined in Eq.~(\ref{M_inv}).  

\subsection{Measurement of the Green's function, density and double occupancy}

The Green's functions can be obtained from the matrix $M$ in a procedure analogous to the one proposed for imaginary-time simulations in Ref.~[\onlinecite{Werner06}]. Particularly simple is the calculation of the density and double occupancy. From the segment representation it immediately follows that $n_\sigma(t)$ is the probability to have a segment of spin $\sigma$ present at time $t$, while $n_\uparrow n_\downarrow(t)$ is the probability to find overlapping segments at time $t$ (taking into account the signs of the Monte Carlo configurations):
\begin{eqnarray}
n_\sigma(t) &=& \frac{\langle \phi_c \delta(\text{segment of type } \sigma \text{ at } t) \rangle_{|w_c|}}{\langle \phi_c \rangle_{|w_c|}},\\
n_\uparrow n_\downarrow (t) &=& \frac{\langle \phi_c \delta(\text{segments of type } \uparrow \text{ and } \downarrow \text{ at } t) \rangle_{|w_c|}}{\langle \phi_c \rangle_{|w_c|}}.
\end{eqnarray}

\subsection{Current measurement}

The current $I_{L\sigma}=-2\text{Im} \sum_{p\in L} V^L_{p,\sigma} \langle a^{L\dagger}_{p,\sigma}d_\sigma \rangle=-2\text{Im}\langle \tilde a^\dagger_{L,\sigma}d\rangle$ can be measured as explained in Ref.~[\onlinecite{Muehlbacher08}]. We expand the quantity
\begin{align}
I_{L\sigma}(t)&=-2\text{Im} Tr \rho_0 \Big(\tilde T e^{i\int_0^t ds H_\text{mix}(s)}\Big)e^{itH_\text{loc}} \tilde a^\dagger_{L,\sigma} d_\sigma e^{-itH_\text{loc}} \Big(T e^{-i\int_0^t ds H_\text{mix}(s)}\Big)
\label{current_hyb}
\end{align}
in powers of $H_\text{mix}$, which leads to the same collection of diagrams as discussed above, except that there is now an operator $d_\sigma$ fixed at time $t$ and that the hybridization functions $\Delta$ connecting to this operator have only an $L$-component:
\begin{equation}
M_\sigma^{-1}(i,j)=i\Delta_L(t^\sigma_{K,i},t'^\sigma_{K,j})+i\Delta_R(t^\sigma_{K,i},t'^\sigma_{K,j})(1-\delta_{t,t^\sigma_{K,i}}).
\end{equation}
Having identified the Monte Carlo configurations $c$ (illustrated in Fig.~\ref{contour_current}) and their weights $w_c$ we can implement a random walk based on $|w_c|$ and measure the current as
\begin{equation}
I_{L\sigma}=\sum_c w_c
=\langle \phi_c\rangle_{|w_c|}\sum_c|w_c|.
\end{equation}
In contrast to the density measurement (which was based on an expansion of the identity so that $\sum_c |w_c|=1/\langle \phi_c\rangle_{|w_c|}$) we cannot directly measure the normalization factor $\sum_c |w_c|$. One possibility to get rid of this unknown factor is to consider the ratio $I/I^{(1)}$ between the current and the lowest order contribution $I^{(1)}$ which can be calculated analytically. Since 
\begin{equation}
I_{L\sigma}^{(1)}
=\langle \phi_c\delta(c \text{ 1st order}) \rangle_{|w_c|}\sum_c|w_c|
\end{equation}
we can measure the current as
\begin{equation}
I_{L\sigma}=I_{L\sigma}^{(1)}\frac{\langle \phi_c \rangle_{|w_c|}}{\langle \phi_c \delta(c \text{ 1st order})\rangle_{|w_c|}}.
\end{equation}

\section{Results: perturbation order, density and double occupancy \label{results}}

\subsection{Perturbation order and average sign}

The average perturbation order increases linearly with the time interval to be simulated, and is not per se the important limiting factor in the simulations. The main constraint is a dynamical sign problem: the factors of $(\pm i)$ associated with each order of the expansion and the complex determinants mean that the average sign of the diagrams contributing to any quantity decays exponentially as the perturbation order is increased .
These phenomena are illustrated in Fig.~\ref{order_t} which presents results obtained using the hybridization expansion algorithm on a model of spinless fermions. The same behavior is found in interacting models and in the weak coupling algorithm. 
The left panel shows the distribution of perturbation orders for simulations over different time intervals. The mean perturbation order can be estimated from the positions of the maxima in these curves. 
The right panel shows the average sign, which is seen to decay exponentially with the length of the time interval to be simulated.
Note that both diagrammatic algorithms can treat temperature $T=0$ without particular difficulties. 

Accurate measurements of physical quantities can be obtained for $\langle \text{sign} \rangle \gtrsim 0.001$, and whether steady state can be reached depends on the method, the parameters, and the observable. Non-zero temperature and voltage bias tend to reduce the sign problem, but not enough to enable simulations on significantly longer contours. The important effect of a non-vanishing voltage bias is to accelerate the convergence to steady state, at least in the weak-coupling approach. As can be seen e.g. from the right hand panel of Fig.~\ref{order_t}, while the basic scaling behavior is the same for all models and parameters, prefactors can depend substantially on details. A careful effort to optimize parameters has not yet been undertaken, but seems likely to be worthwhile.

\begin{figure}[t]
\begin{center}
\includegraphics[angle=-90, width=0.49\columnwidth]{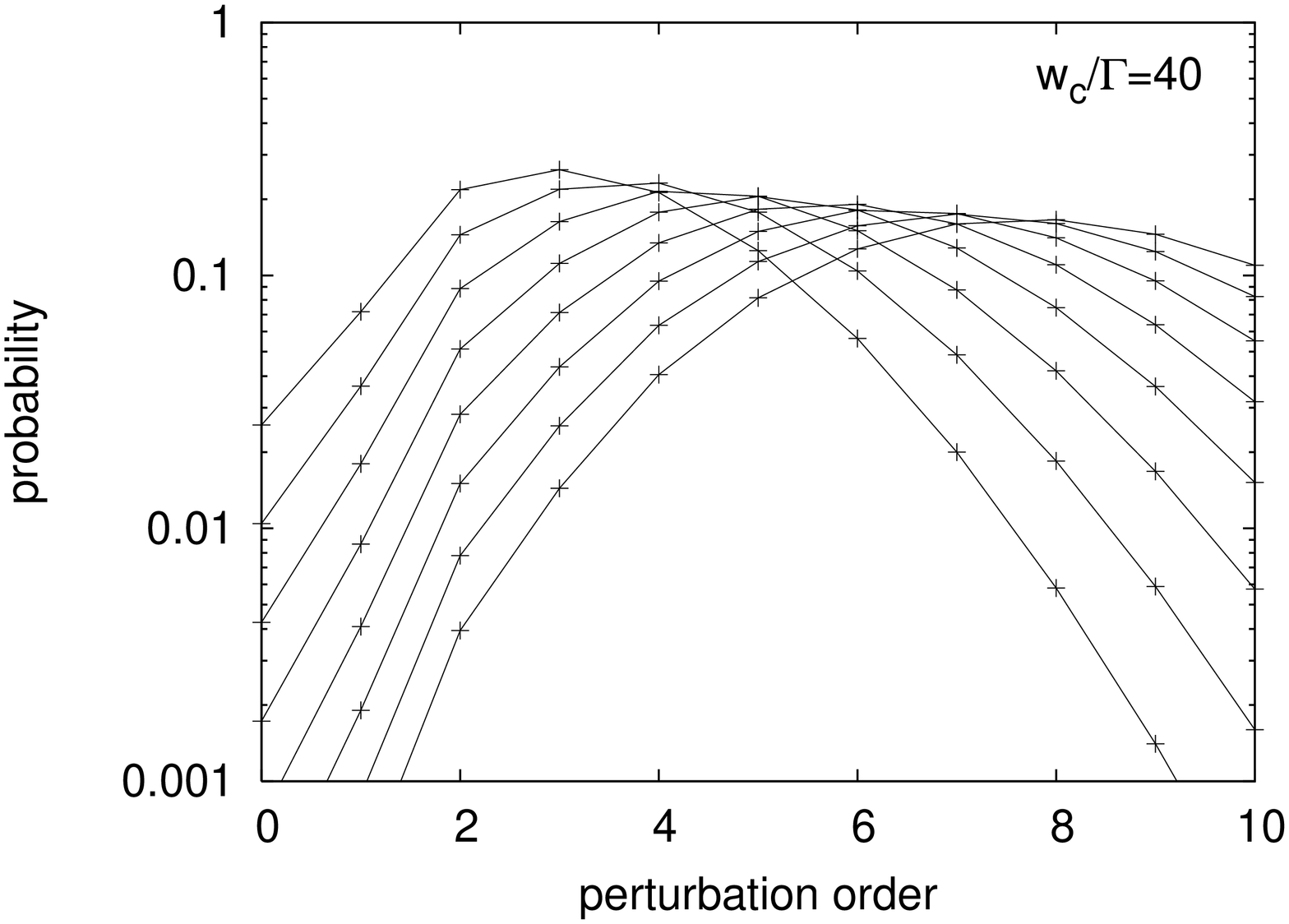}
\includegraphics[angle=-90, width=0.49\columnwidth]{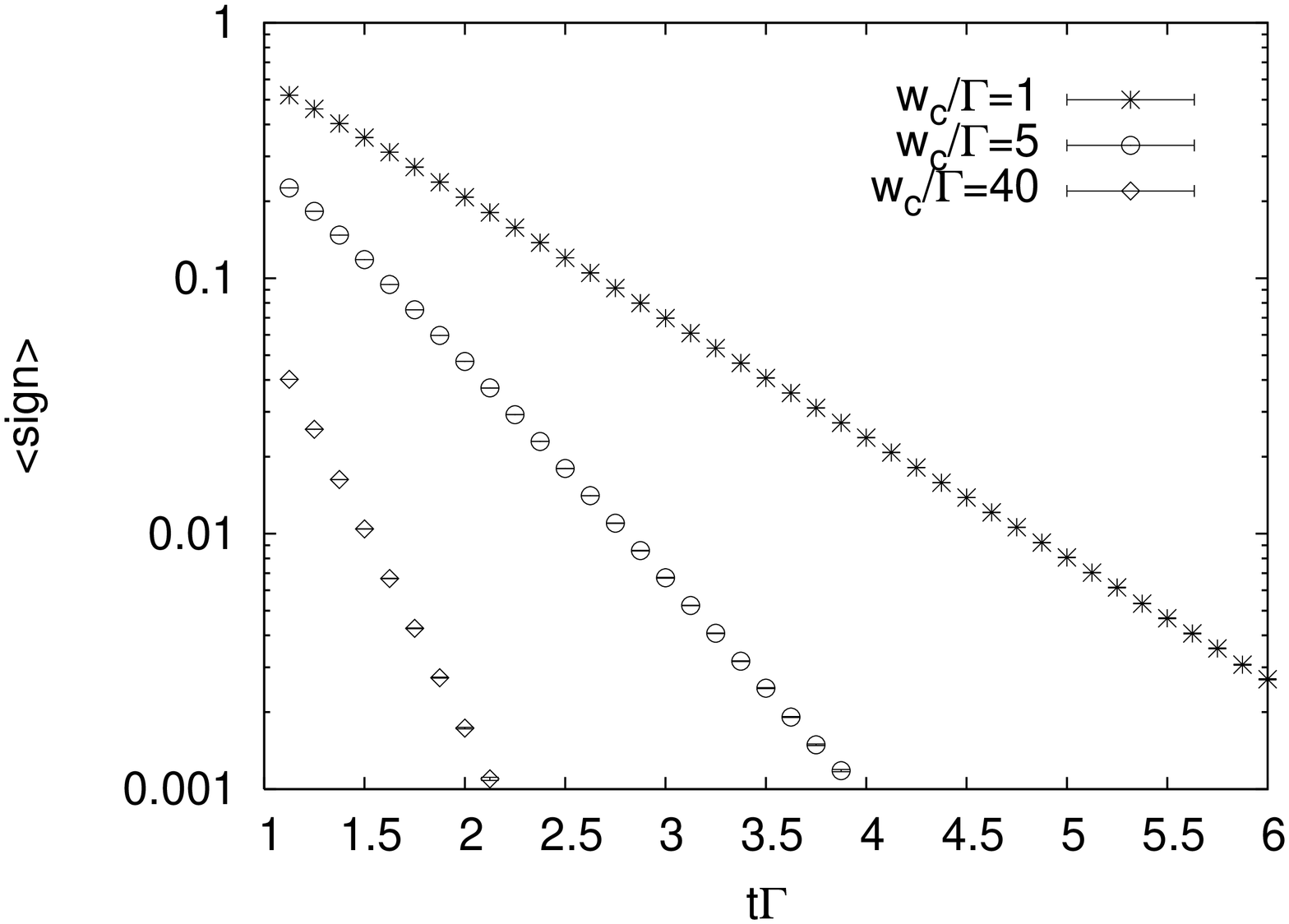}
\caption{Distribution of perturbation orders and average sign obtained using the hybridization expansion algorithm for a non-interacting dot with soft cutoff, $T=0$, $V=0$, $\epsilon_d/\Gamma=-0.5$ and a single species of fermion. Left panel: distribution of perturbation orders for different lengths of the contour ($t\Gamma=1.25, 1.50, \ldots, 3$ from left to right) and cutoff $\omega_c/\Gamma=40$. The average perturbation order grows $\sim t$. Right panel: average sign as a function of time for indicated values of the cutoff. 
}
\label{order_t}
\end{center}
\end{figure}

The left panel of Fig.~\ref{order_u} shows that in the weak-coupling approach, the average perturbation order (at fixed $t$) depends on the interaction strength.  
As in the imaginary-time version of this algorithm,\cite{Gull08} the perturbation order grows roughly linearly with increasing $U$, making it difficult to study dots with $U/\Gamma\gtrsim 3$. Larger values of $K$ also lead to a larger perturbation order and hence to a more severe sign problem, as illustrated in Fig.~\ref{order}. 
The parameter $K$ can also be chosen negative or complex (we used $K=-0.01$ in all our weak-coupling simulations).
In this case, $\gamma$ is complex and some of the phase oscillations are shifted from the $(\pm i K)^{n_{\pm}}$ to the determinant. It is in principle possible to choose different constants $K_+$ and $K_{-}$ on the forward and backward contour. If $K_+=iK=-K_{-}$ (with $K$ positive), then all the phase oscillations from the $(\pm iK_{\pm})^{n_{\pm}}$ are eliminated. However, it turns out that $K_+=-K_{-}$ leads to a perturbation order which is about the same as for $iK$ on both the forward and backward contour, which in turn is somewhat worse than $-K$ on both the forward and backward contour, so that there
appears to be
no particular advantage in this choice of $K$. 

\begin{figure}[t]
\begin{center}
\includegraphics[angle=-90, width=0.49\columnwidth]{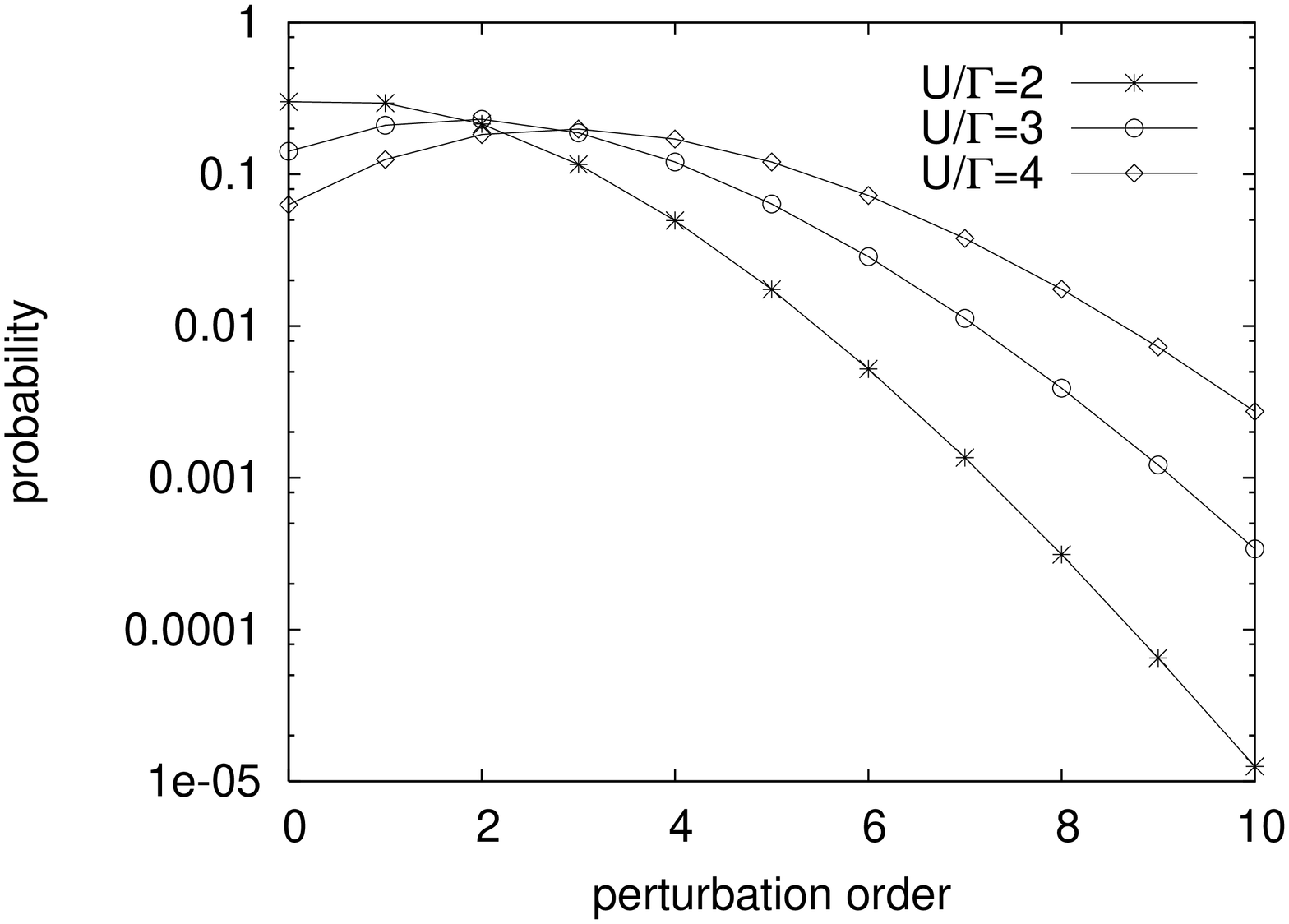}
\includegraphics[angle=-90, width=0.49\columnwidth]{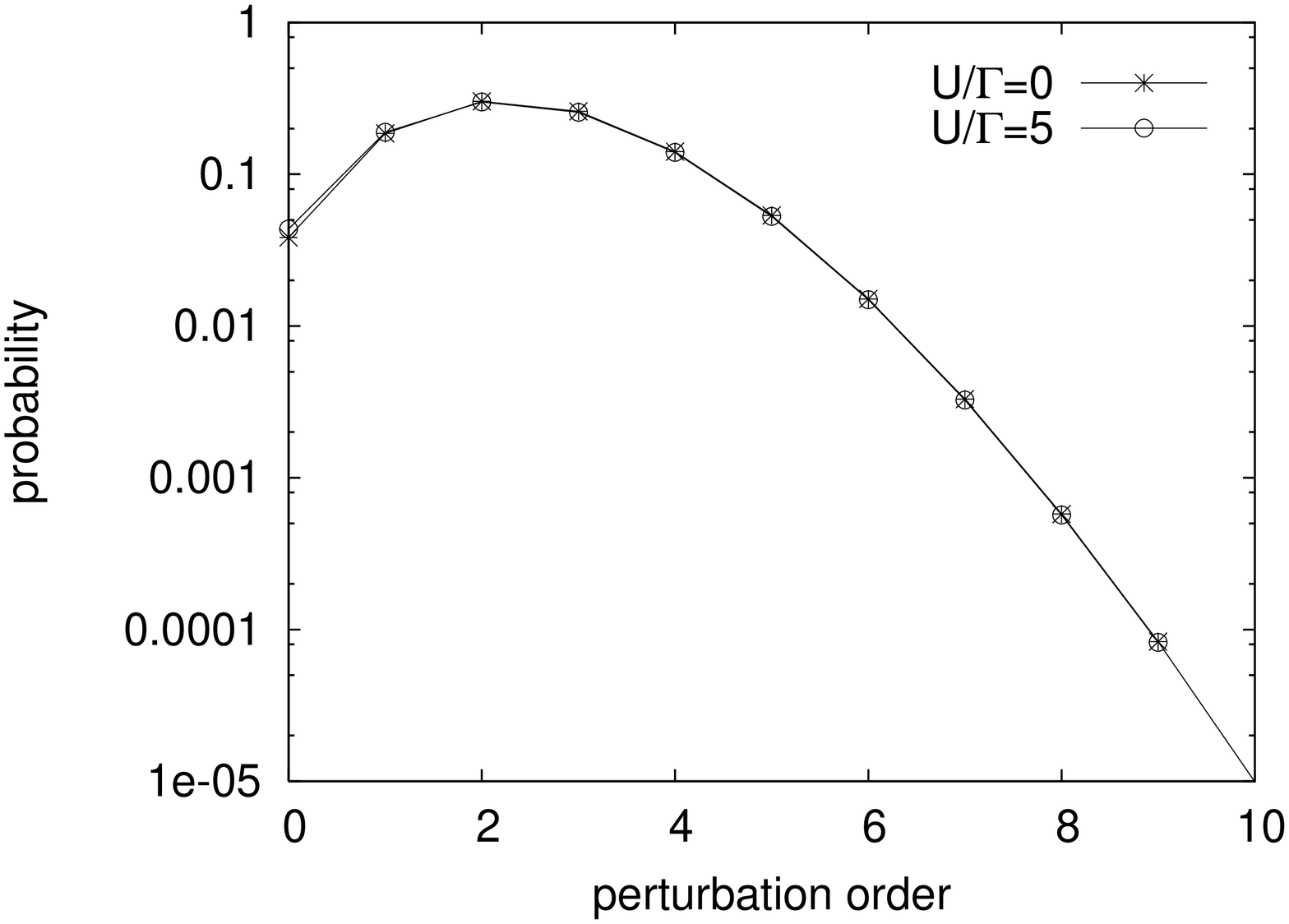}
\caption{Distribution of perturbation orders for different values of the interaction $U$, $T=0$, $V=0$, $\epsilon_d+U/2=0$. The left panel shows the distribution of perturbation orders for the weak-coupling algorithm ($t\Gamma=1$, infinite bandwidth), where the average perturbation order grows $\sim U$. 
Right panel: distribution of perturbation orders for the hybridization expansion algorithm ($t\Gamma=1.5$, $\omega_c/\Gamma=10)$. Here there is almost no dependence on interaction strength.
}
\label{order_u}
\end{center}
\end{figure}

\begin{figure}[t]
\begin{center}
\includegraphics[angle=-90, width=0.49\columnwidth]{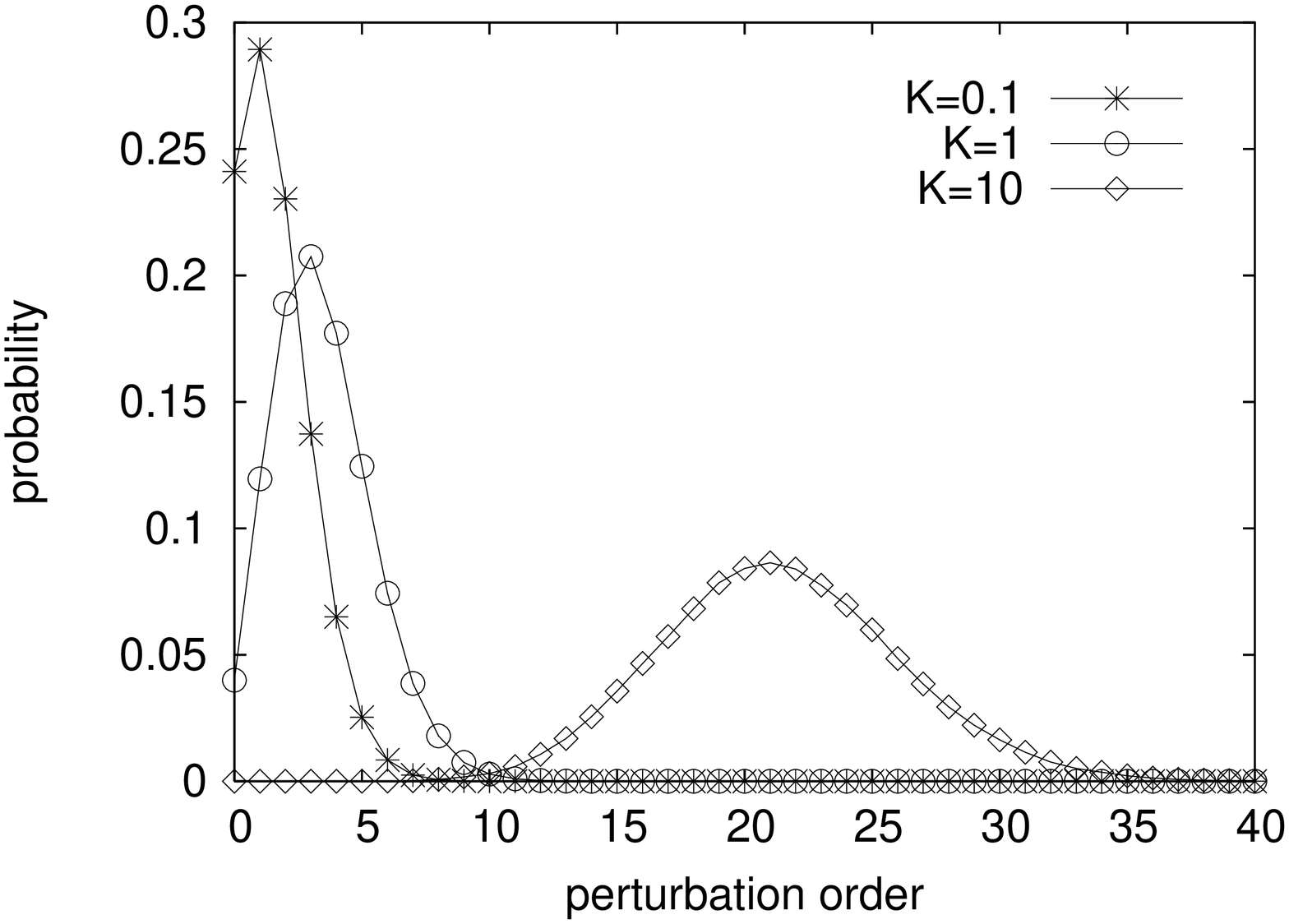}
\includegraphics[angle=-90, width=0.49\columnwidth]{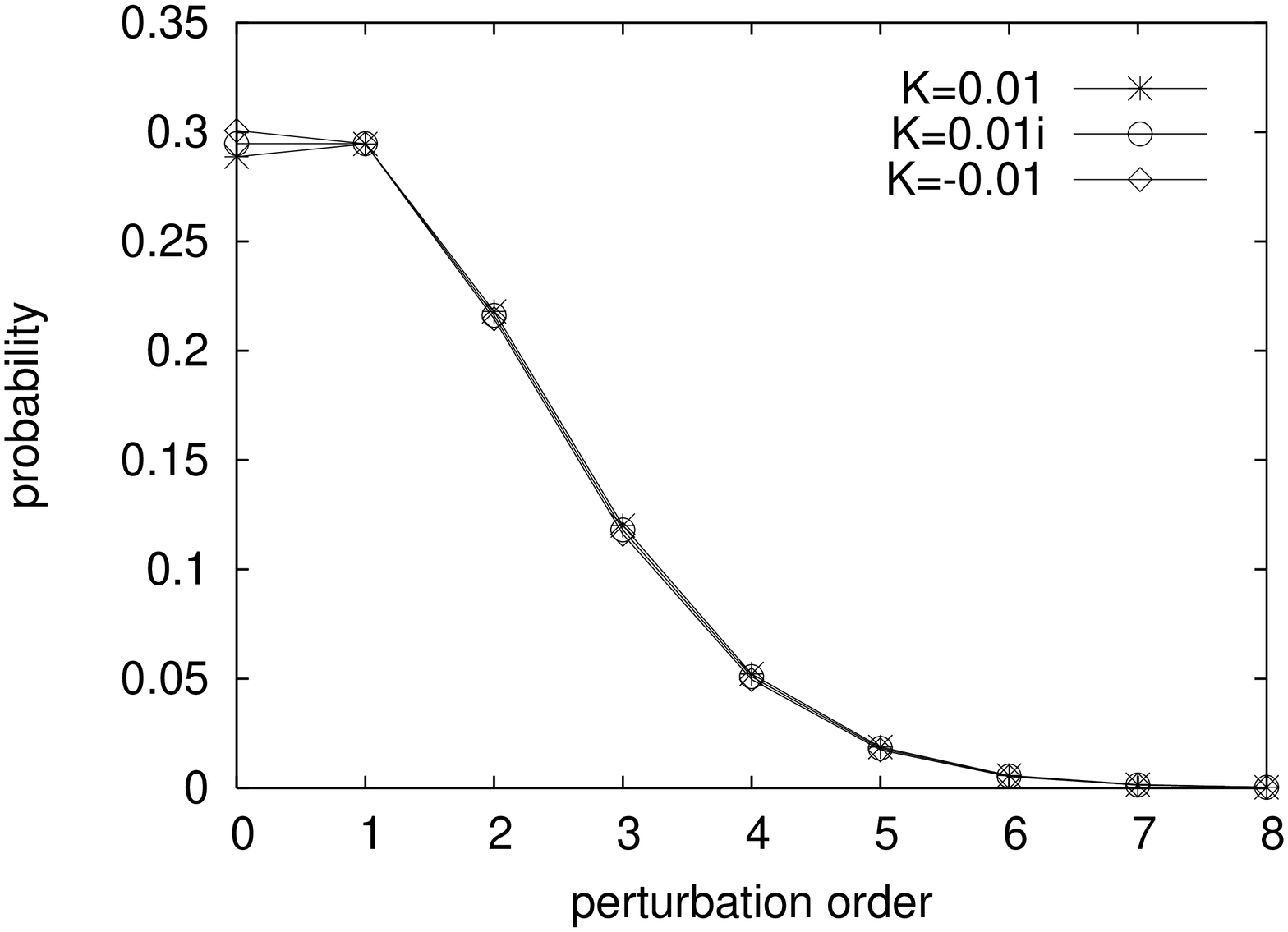}
\caption{Left panel: perturbation order distribution for $t\Gamma=1$, $U/\Gamma=2$, $V/\Gamma=0$ and several positive values of $K$. The average signs in these simulations are 0.24 ($K=0.1$), 0.04 ($K=1$) and 0.0002 ($K=10$). Right panel: complex $K$ of norm 0.01. The best choice appears to be the negative $K$. 
}
\label{order}
\end{center}
\end{figure}

The right panel of Fig.~\ref{order_u} shows that in the hybridization expansion algorithm, the average perturbation order is essentially independent of interaction strength. This is in contrast to the imaginary-time version of this algorithm,\cite{Werner06} where the perturbation order decreases with increasing interaction strength. From Eq.~(\ref{weight_strong}) it follows that the interaction term merely adds a phase to the Monte Carlo weight and therefore does not affect $|w(c)|$.
While the algorithm can treat strong interactions, it is limited to finite bandwidth, since the average perturbation order diverges as the bandwidth goes to infinity (this is the reason for the dependence on cutoff seen in the right 
hand panel of Fig.~\ref{order_t}).
We find, however, that systems with larger cutoff reach steady state more rapidly (as in Fig.~1 of Ref.~[\onlinecite{Schmidt08}]). We have not yet attempted to optimize the cutoff to strike the best balance between perturbation order and time needed to reach steady state. Such an optimization would be worth while. 

\subsection{Density and double occupancy}

\begin{figure}[t]
\begin{center}
\includegraphics[angle=0, width=0.49\columnwidth]{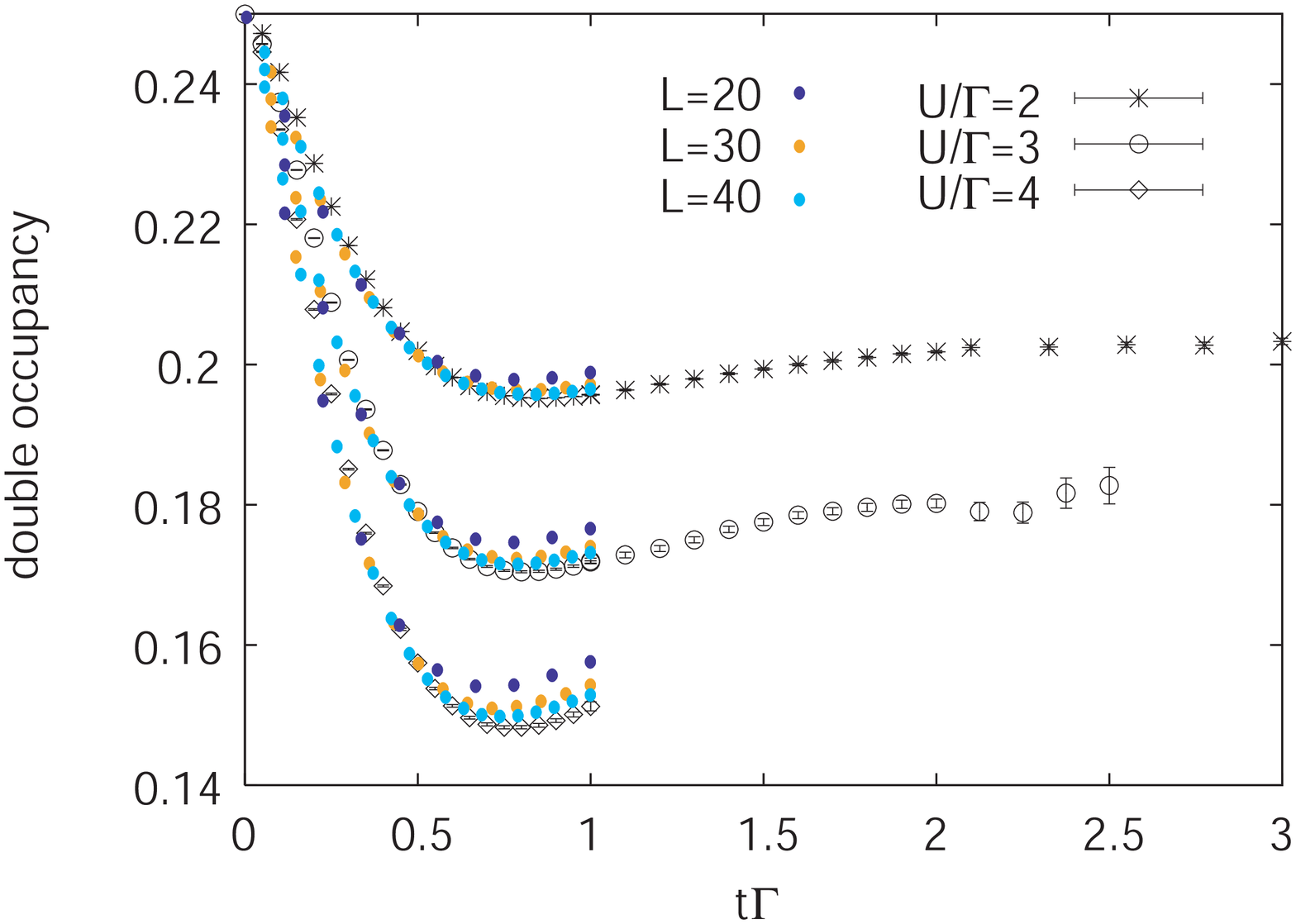}
\includegraphics[angle=0, width=0.49\columnwidth]{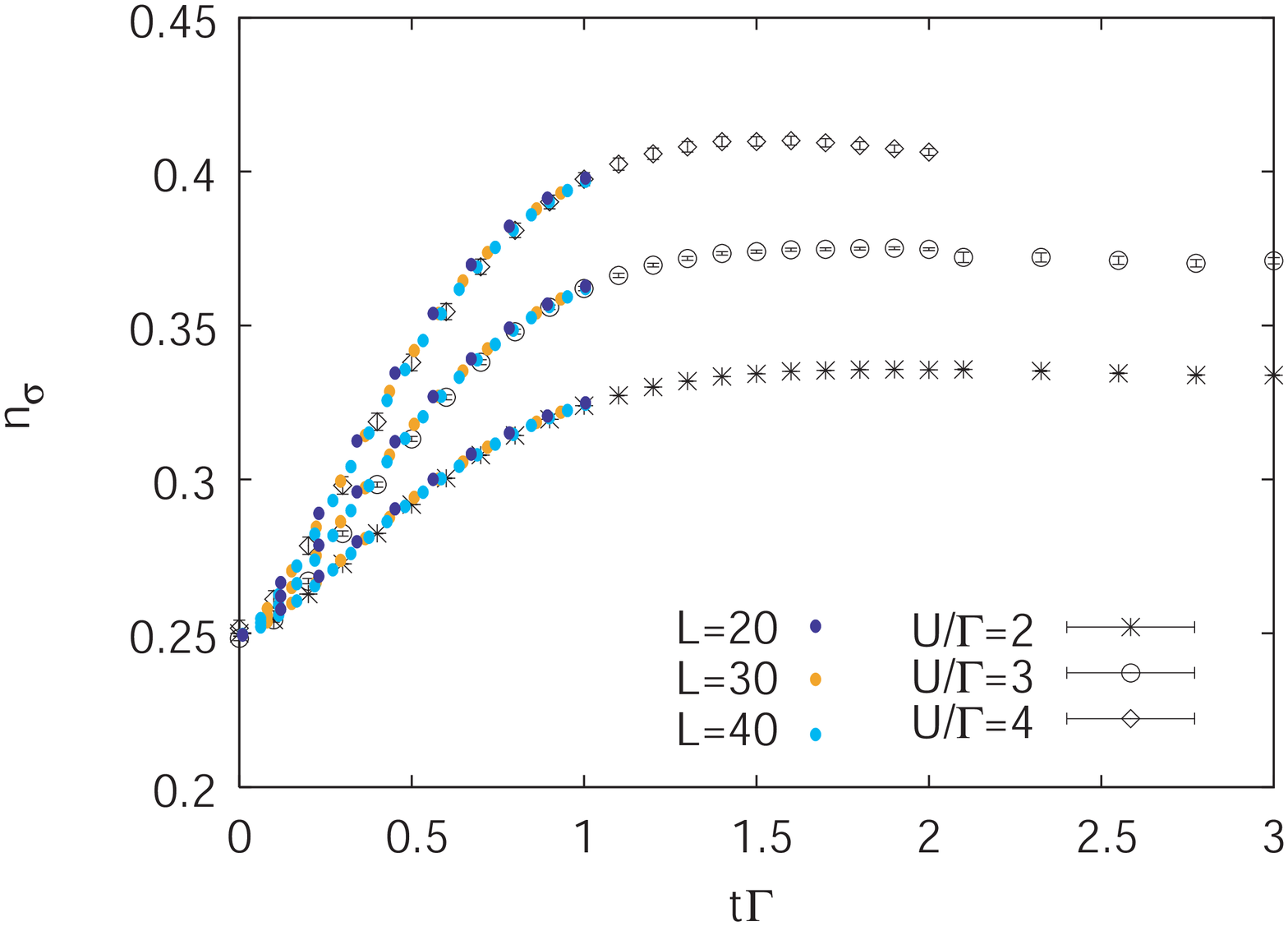}
\caption{(color online) Weak-coupling results for the double occupancy and density computed at $T=0$ and $V=0$ with parameter $K=-0.01$ and for infinite bandwidth. Left panel: double occupancy for $U/\Gamma=2, 3, 4$ at half filling ($\epsilon_d+U/2=0$). Right panel: density per spin for $\epsilon_d+U/2=\Gamma$ with the other parameters the same. 
The full dots show the results from the real-time Hirsch-Fye method for $t\Gamma=1$ and indicated values of the number of time slices $L$.
}
\label{double_weak}
\end{center}
\end{figure}

The left panel of Fig.~\ref{double_weak} shows as black symbols the evolution of the double occupancy obtained using the weak-coupling algorithm in equilibrium ($V=0$) and at zero temperature for a system tuned to be at half filling. At time $t=0$ the double occupancy takes the value $0.25$ appropriate to the noninteracting half filled system. The effect of the interactions is to reduce it. 
The right panel shows the dot occupancy per spin, computed for a level position corresponding to a quarter-filled dot at $U=0$. Turning on the interaction increases the dot occupancy; this is a precursor of the Coulomb blockade plateau. 

One sees from the figure that for $U/\Gamma=2$ it is possible to obtain a good estimate of the steady state value, whereas for $U/\Gamma=4$ the perturbation order grows too rapidly with $t$ and the sign problem becomes severe before the system approaches the steady state. Whether or not steady state can be reached depends on the observable. The statistics for the density is significantly better than for the double occupancy, so density calculations can be carried to longer times.

We also show in Fig.~\ref{double_weak} results obtained with the real-time Hirsch-Fye method for $t\Gamma=1$ and different numbers of time slices. As the number of time slices is increased, the systematic error due to the Trotter break-up decreases and the result approaches the continuous-time curves (which are free of systematic errors). Comparison of the left and right panel shows that the density is less sensitive to Trotter errors than the double-occupancy. 
Because the sign problem in the real-time Hirsch-Fye method becomes severe for $L\gtrsim 30$, longer times can only be reached at the expense of larger discretization errors. 
In our calculations we found that the weak-coupling continuous-time algorithm allows to roughly double the time interval which can be simulated, compared to Hirsch-Fye.

Simulations at $V=0$ and $T=0$ suffer from the most severe sign problem. At non-zero voltage bias, the system reaches steady state more rapidly, as illustrated in the left hand panel of Fig.~\ref{double_v}. For $U/\Gamma=2$ and 3 we can therefore obtain an accurate estimate of the steady state double occupancy. The voltage dependence of this quantity is plotted in the right hand panel of Fig.~\ref{double_v}. As the voltage bias is increased, the steady state double occupancy drops, reaches a minimum and then increases with increasing $V$ toward the non-interacting value of $0.25$. The initial drop in the double occupancy is the result of the destruction of Fermi liquid coherence with increasing voltage bias, similar to the destruction caused in equilibrium by a non-zero temperature. At larger voltage bias a reversion towards the non-interacting value of $0.25$ is evident. This non-monotonic behavior was also observed by the time-dependent density matrix renormalization group method.\cite{Kirino08}

\begin{figure}[t]
\begin{center}
\includegraphics[angle=-90, width=0.49\columnwidth]{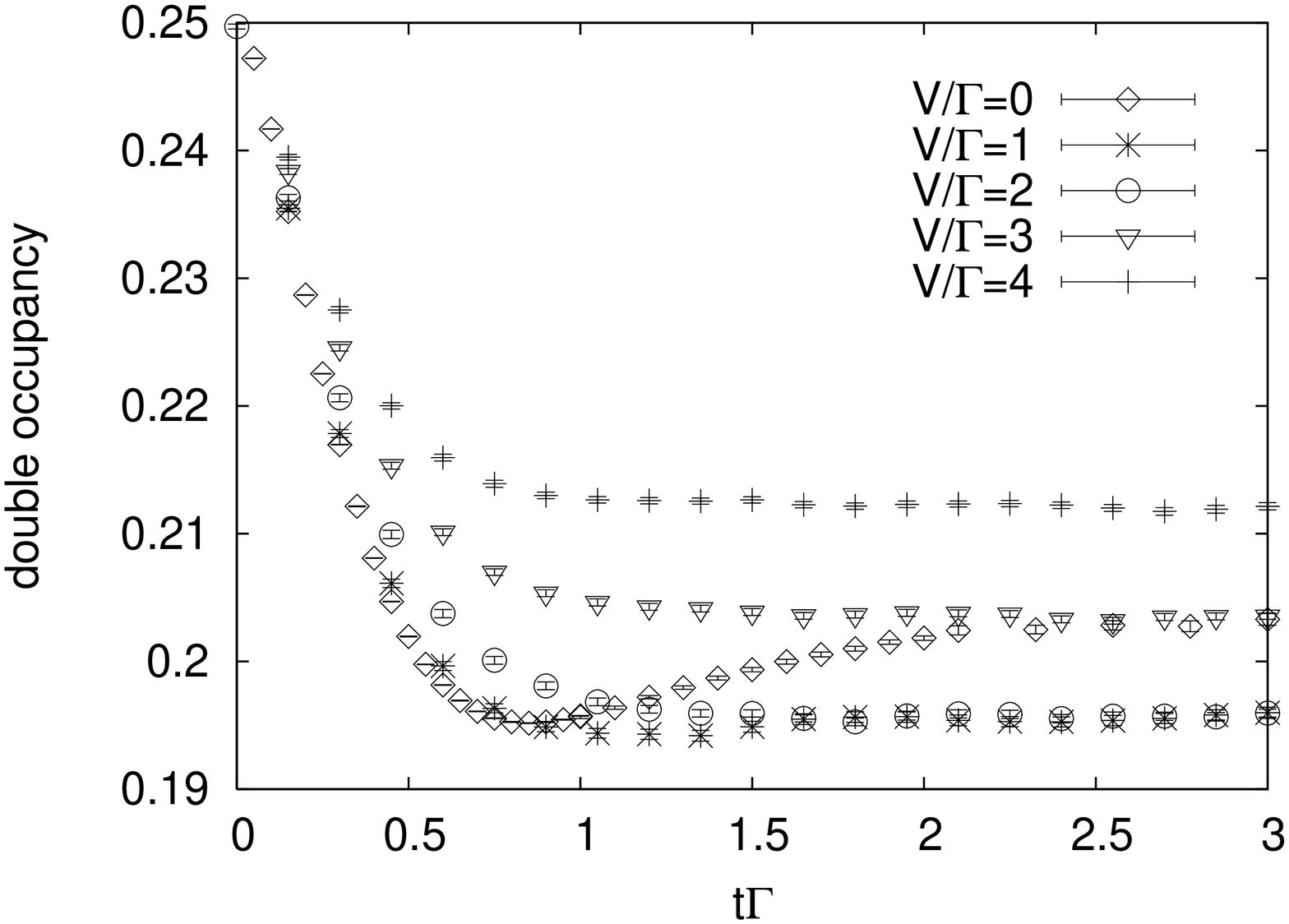}
\includegraphics[angle=-90, width=0.49\columnwidth]{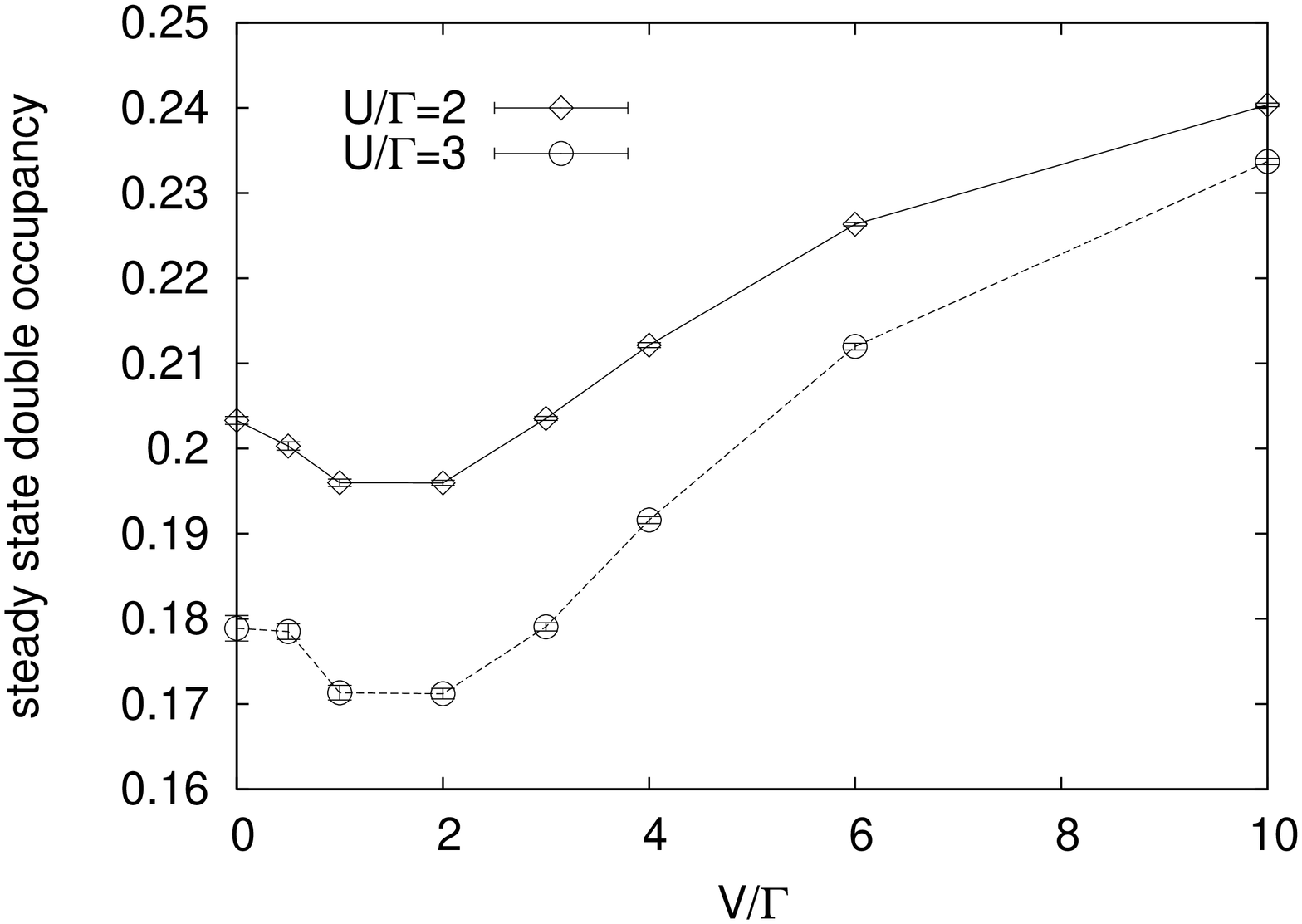}
\caption{Voltage dependence of the double occupancy for an infinite flat band, $T=0$, half filling ($\epsilon_d+U/2=0$). The left panel shows the time evolution of the double occupancy for $U/\Gamma=2$ and indicated values of $V$. At finite voltage bias, the system reaches steady state much more rapidly than for $V=0$. Right panel: steady state value of the double occupancy as a function of voltage bias for $U/\Gamma=2$ and 3, $T=0$ and half filling.}
\label{double_v}
\end{center}
\end{figure}

Figure~\ref{double_strong} shows results for double occupancy and dot occupation obtained using  the hybridization expansion algorithm. Here, the initial state is an empty dot which is decoupled from the leads and at $t=0$ we turn on the hybridization.  The left panel shows the time evolution of the double occupancy in a dot with $\epsilon_d+U/2=0$, a hard cutoff $\omega_c/\Gamma=10$, $\beta \Gamma=\nu\Gamma=10$, and the right panel shows the evolution of the density per spin. Increasing the interaction accelerates the approach to equilibrium, and leads to a slight overshooting of $n(t)$. The reduction of the double occupancy is roughly consistent with the result from the weak-coupling simulation (note that the band widths are different). 
\begin{figure}[t]
\begin{center}
\includegraphics[angle=-90, width=0.49\columnwidth]{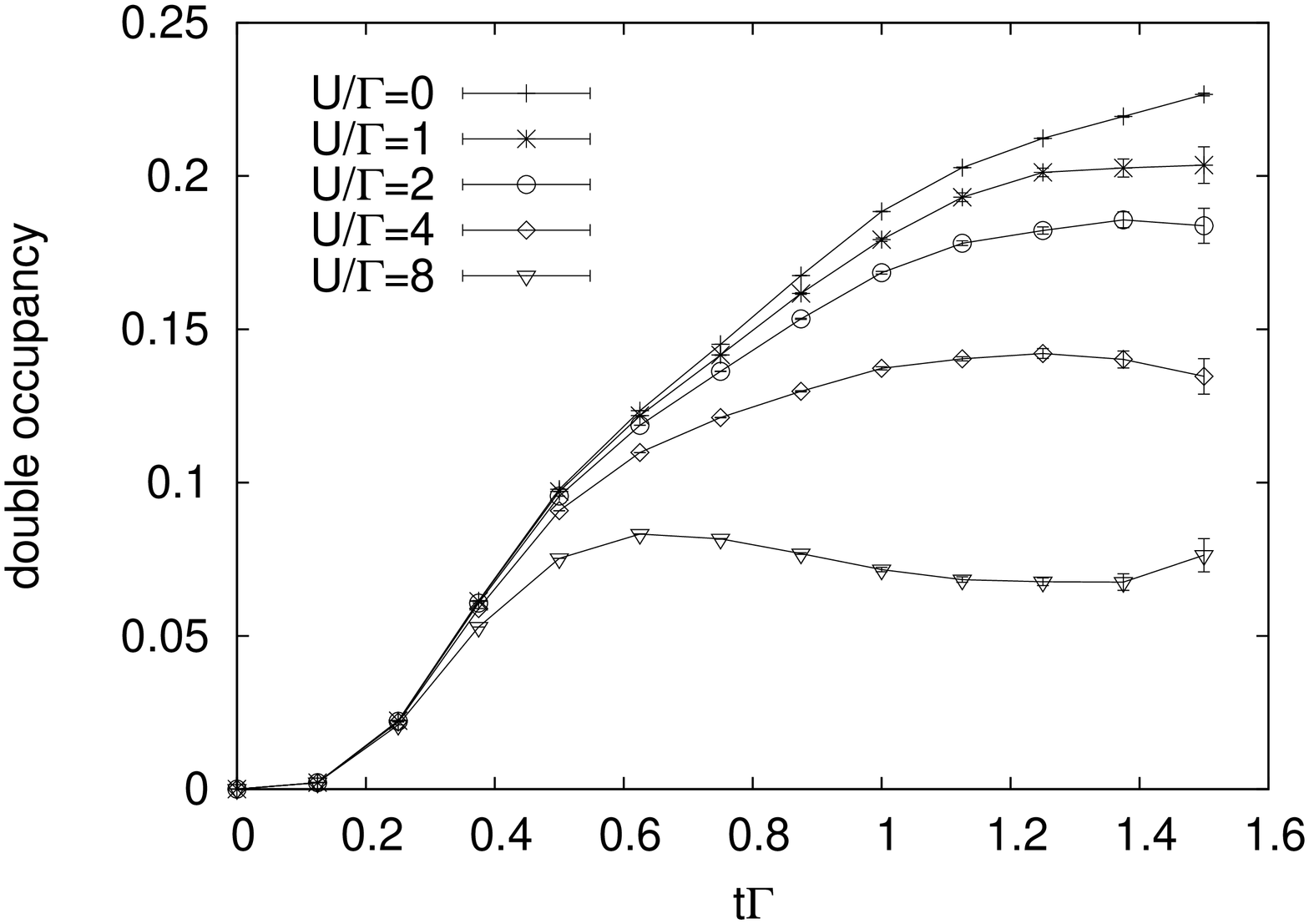}
\includegraphics[angle=-90, width=0.49\columnwidth]{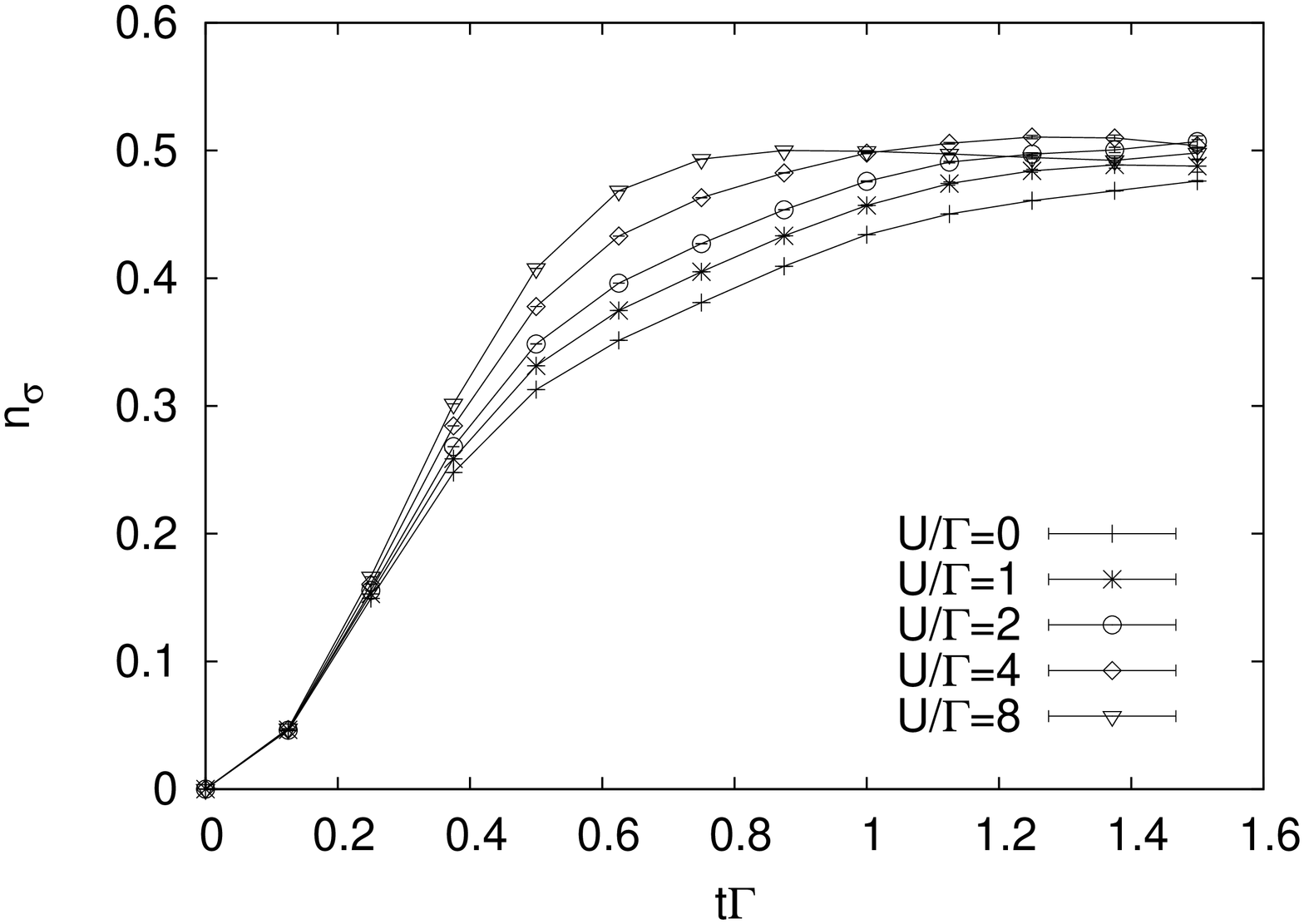}
\caption{Hybridization expansion results for the double occupancy and density. Left panel: double occupancy for indicated values of $U$, 
band cutoff $\omega_c/\Gamma=10$, $\beta \Gamma=\nu\Gamma=10$, $V=0$, $\epsilon_d+U/2=0$. Right panel: density per spin for the same parameters. The initial state is an empty dot decoupled from the leads and at $t=0$ the dot-lead hybridization is turned on. 
}
\label{double_strong}
\end{center}
\end{figure}

\section{Results: Current \label{current}}

\subsection{Qualitative Picture: perturbative and mean field results}

\begin{figure}[t]
\begin{center}
\includegraphics[angle=0, width=0.4\columnwidth]{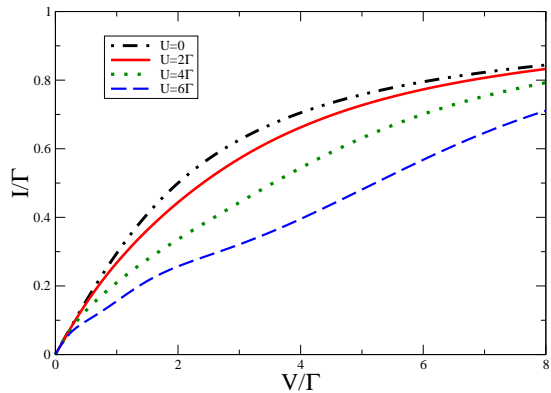}
\caption{Current computed as a function of voltage bias in the infinite bandwidth model using $4^{th}$ order perturbation theory in the interaction $U$ (from Ref.~[\onlinecite{Fuji03}]).}
\label{perturbation}
\end{center}
\end{figure}

To orient the discussion of our results
for the current
we present here a brief outline of the expected qualitative behavior, along with perturbative and mean-field calculations.  In the noninteracting limit, the $d$-density of states of the model defined by Eq.~(\ref{HQI}) takes an approximately Lorentzian form with a peak at the $d$-level energy $\varepsilon_d$ and a width of order $\Gamma$. As $U$ is increased, the structure of the $d$-density of states changes: the peak broadens, and at large enough $U$ splits into two.  The density of states in the region between the two peaks becomes small, except that  in the $V\rightarrow0$, $T\rightarrow 0$ limit a narrow peak (the Kondo resonance) appears at the Fermi level, so the Fermi level density of states remains essentially unrenormalized.
At temperatures or voltage biases greater than the Kondo scale (which becomes exponentially small at strong couplings) the Kondo peak is believed to be destroyed, leaving only the small density of states (Coulomb blockade) behavior.

The current $I$ is, up to various constants, given by the  integral over the voltage window $-V/2<\varepsilon<V/2$ of the product of  $\Gamma_L\Gamma_R/\Gamma^2$ and the density of states. In the noninteracting case $I$ starts out linearly with $V$ and saturates for $V\gg\Gamma$.  As $U$ is increased the broadening of the peak means that the $V$-value needed to reach current saturation increases. The Kondo physics implies that the $T=0$ linear response current is essentially independent of interactions, but what happens at larger $V$ in the strongly correlated regime is unclear.

Figure~\ref{perturbation} shows the current computed from $4^{th}$ order perturbation theory in $U$ 
by Fujii and Ueda \cite{Fuji03} for the infinite bandwidth version of Eq.~(\ref{HQI}). The initial linear rise and eventual saturation of the current are clearly visible, as is the increase in the saturation voltage as $U$ is increased. Also visible in the calculation is the $U$-independence of the linear-response current and hints of the formation of the Coulomb blockade plateau at intermediate $V$ and larger $U$. Of course, the reliability of low-order perturbation theory at these interaction strengths may be questioned.

Figure~\ref{meanfield} shows the results of computations performed using  mean field theory \cite{Komnik04} as well as phenomenological generalizations. Details of the calculations  are given in the Appendix, but the essence is as follows.  In mean field theory of the model studied here, at $T=0$ and in equilibrium,  a transition occurs at $U_c=\pi\Gamma$ between an unpolarized weak coupling state and a strong coupling state characterized by a frozen local moment and a spectral function split into upper and lower Hubbard bands.  This is the mean field representation of Coulomb blockade. The mean field theory does not capture the Kondo effect, so for $U>U_c$ the near Fermi surface density of states is simply suppressed. For $U>U_c$, as the voltage is increased, the degree of spin polarization decreases and within mean field theory we find a sharp phase transition at which the properties revert to those of the unpolarized state. The dashed-dotted curve in the inset of Fig.~\ref{meanfield} shows that for the model studied the transition is first order (jump in $m$) and  for $U=12\Gamma$ occurs at $V \approx 7\Gamma$.  Ref.~[\onlinecite{Komnik04}] presented analytical arguments that a polarized phase would extend to infinite voltage, at least in a model with an infinite bandwidth; this behavior is not found in our numerical solution of the finite-bandwidth model.  
The main panel of Fig.~\ref{meanfield} shows the mean field current computed in various approximations. The dashed double-dotted trace (black on-line) shows the current at $U=0$ (the small differences from the $U=0$ trace in Fig.~\ref{perturbation} arise because in Fig.~\ref{meanfield} a finite bandwidth is used, whereas in Fig.~\ref{perturbation} an infinite-bandwidth limit is taken). The dashed-dotted curve (red on-line) shows the current computed from mean field theory. Comparison to the noninteracting (dashed double-dotted) curve reveals the Coulomb-blockade suppression of the current at small bias and the reversion to the noninteracting result at higher bias. Within mean field theory the reversion occurs via a first order transition at a critical bias 
of the order of one half of the Coulomb gap.
Mean field theory is of course not an entirely accurate description. For example, as noted by the authors of Ref.~[\onlinecite{Komnik04}], the transition is an artifact of mean field theory.  One would expect features of the Coulomb gap to persist at high voltage biases. To qualitatively assess the consequences of this physics we show as the solid line (blue on-line) the results of a computation in which the Coulomb gap (splitting of the density of states into two peaks) is fixed at its $V=0$ value. A broader range of current suppression and a high saturation voltage are evident. 

\begin{figure}[t]
\begin{center}
\includegraphics[angle=0, width=0.45\columnwidth]{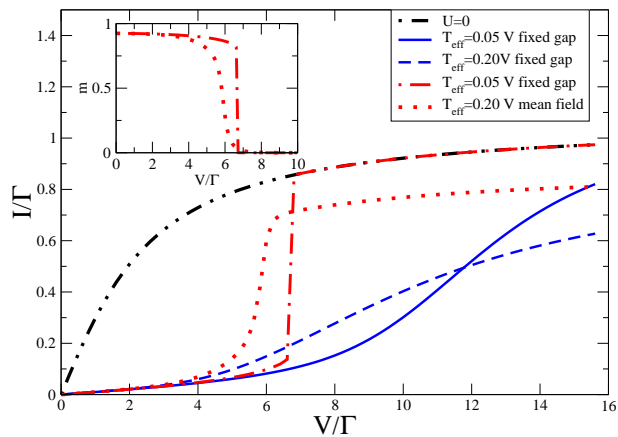}
\caption{Main panel: dashed-dotted line (red on-line): current computed in mean field theory for $U=12\Gamma$ as a function of voltage bias $V$ assuming negligible pseudothermal broadening. Dotted line (red on-line): current computed in mean field theory assuming a pseudothermal  broadening equal to $20\%$ of the voltage bias. Solid line (blue on-line): current computed using mean field theory with gap fixed at the $V=0$ value and negligible pseudothermal broadening; dashed line (blue on-line): current computed using mean field theory with gap fixed at the $V=0$ value and pseudothermal broadening equal to $20\%$ of the voltage bias.  Dash double-dotted line (black on-line): current computed for the non-interacting model (negligible pseudothermal broadening). Inset: dot magnetization as a function of voltage bias $V/\Gamma$  computed in mean field theory for negligible pseudothermal broadening (dot-dashed line, red-on line) and moderate pseudothermal broadening (dotted line, red on-line). All computations were performed for a hard cutoff with $\omega_c=10\Gamma$ and $\nu\Gamma=10$.}
\label{meanfield}
\end{center}
\end{figure}

The mean field theory is deficient in an additional way. Ref.~[\onlinecite{Mitra05}] showed that mean field theory misses the fact that bias  voltage functions as an effective temperature (proportional to the bias voltage times a numerical factor related to scattering phase shifts) which broadens all of the properties. In order to asses the qualitative effect of this consideration we modeled the pseudothermal broadening effect of a voltage bias by performing the calculations at a temperature chosen to be $T_\text{eff}=0.2V$. The dotted lines in the inset and main panel show that the pseudothermal broadening effect converts the first order transition into a second order one. More significantly, we see that including an effective temperature tends to decrease the current at higher biases. The numerical calculations discussed below will be seen to be most consistent with the fixed gap, pseudothermally broadened mean field calculations.

\subsection{QMC results: Current}

\subsubsection{Weak Coupling Expansion}

In the weak-coupling simulations, the current starts from the steady-state value for the non-interacting dot, $-4\text{Im} A(0,0)$, decreases in magnitude after the interactions are turned on, and eventually converges at sufficiently long times at the value corresponding to the steady state current through the interacting dot. As shown in Fig.~\ref{current_weak}, useful estimates for this steady state value can be obtained for interaction strengths $U/\Gamma\lesssim 3$. While the first term in Eq.~(\ref{A}) can be computed directly for the wide band limit, the second integral requires a frequency cutoff $\omega_c$. However, we found that the current results are insensitive to this cutoff-value, as long as $\omega_c  \gtrsim V$. All our results were obtained for $\omega_c/\Gamma=10$.

Figure \ref{current_weak} presents the time dependence of the current obtained from the weak coupling algorithm for different values of the voltage bias at $U=2\Gamma$ (left panel) and at different interaction strengths for fixed bias $V=4\Gamma$ (right panel).  We see that the effect of the interaction is to reduce the magnitude of the current. However, the corrections are relatively small at the $U$ and bias voltages studied: the noninteracting systems already gives a good approximation to the current if the interaction is not too strong.

\begin{figure}[t]
\begin{center}
\includegraphics[angle=-90, width=0.49\columnwidth]{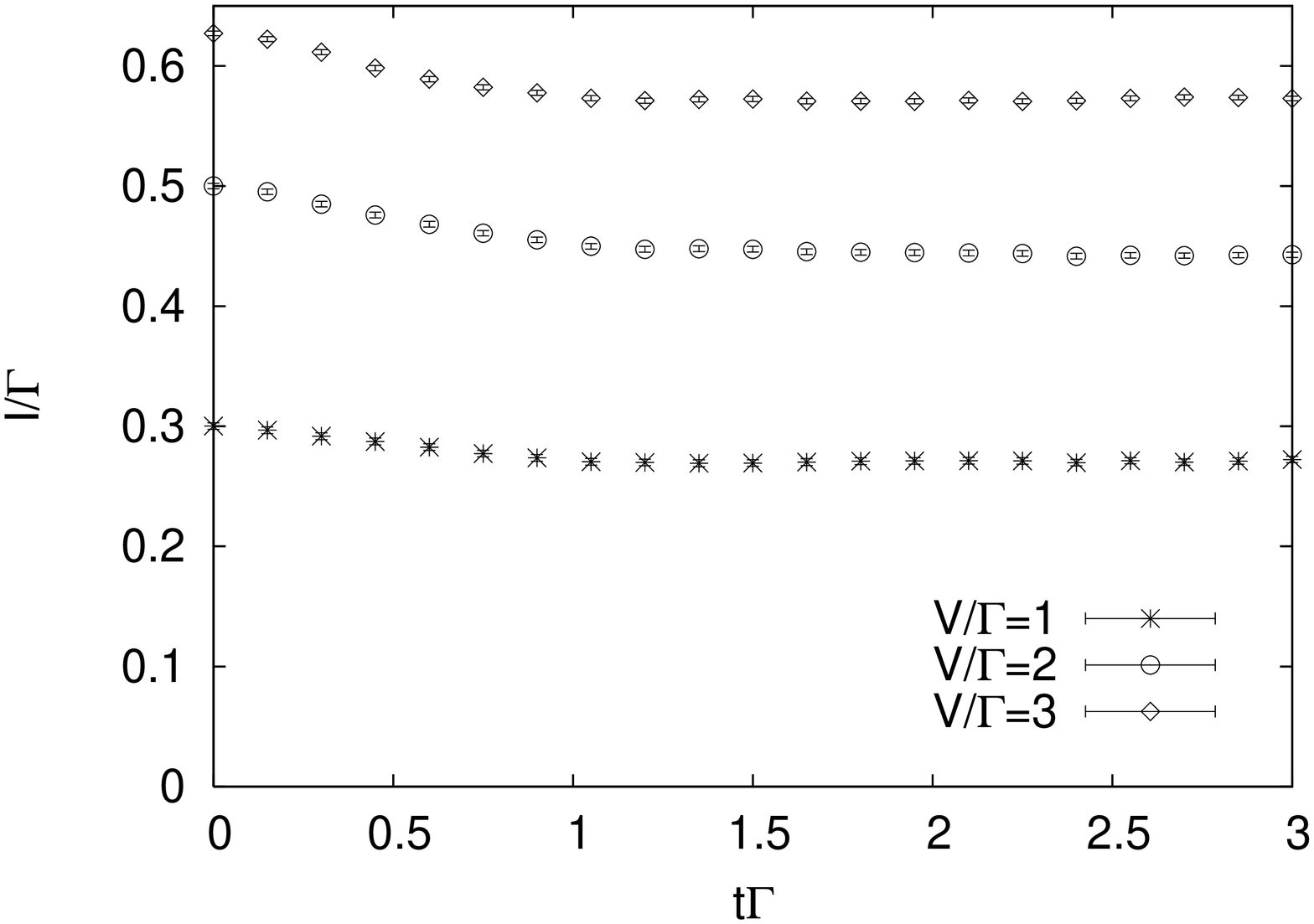}
\includegraphics[angle=-90, width=0.49\columnwidth]{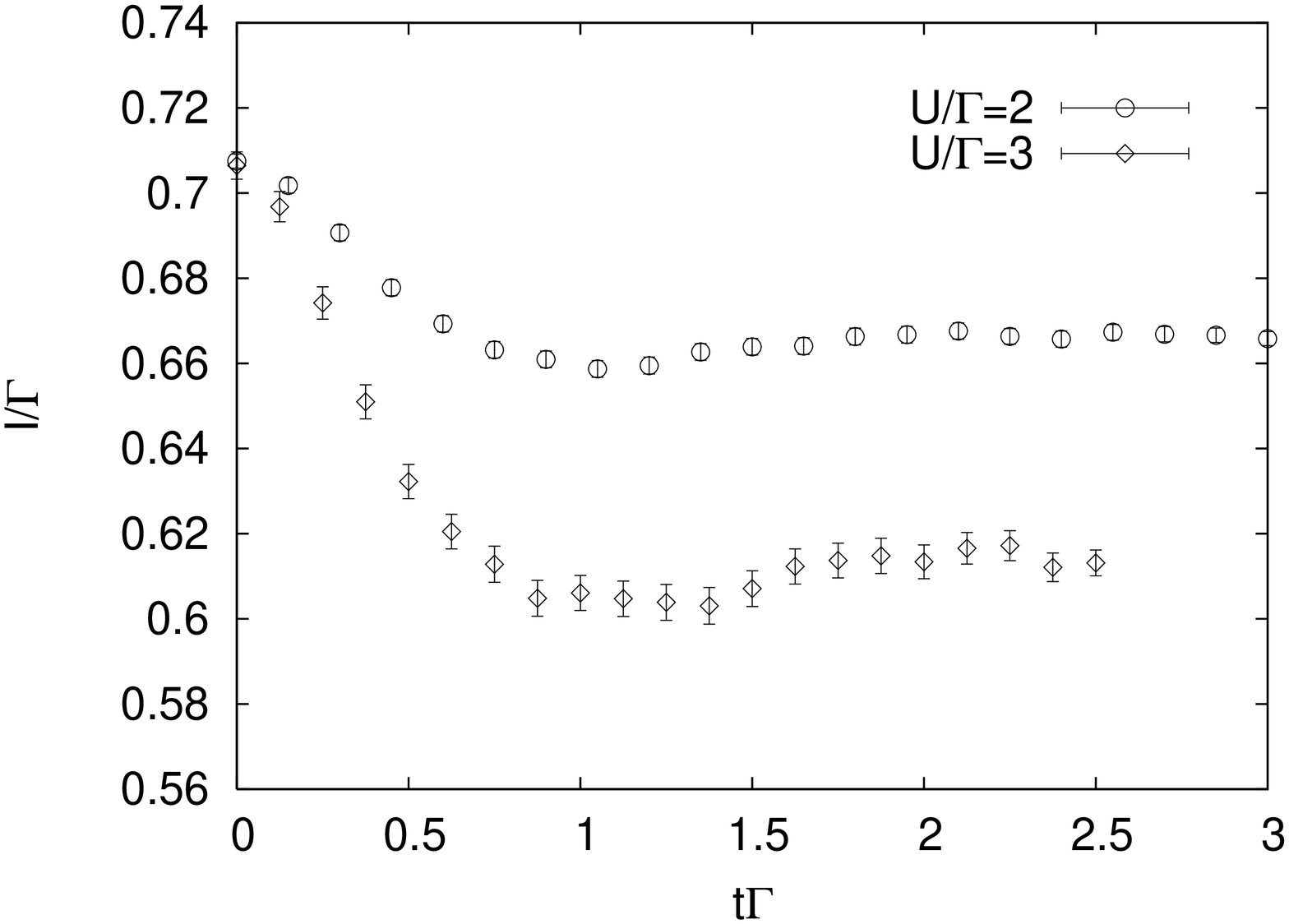}
\caption{Weak-coupling expansion results for the current at temperature $T=0$. The left panel shows the  time evolution for $U/\Gamma=2$ and indicated values of the voltage bias. The value at time $t=0$ corresponds to the non-interacting current $I_0=-4\text{Im}A(0,0)$. As the interaction is turned on, the magnitude of the current decreases and eventually converges to the steady state value for the interacting dot. The right panel shows the time dependence of the current for $V/\Gamma=4$ and $U/\Gamma=2$ and 3. 
}
\label{current_weak}
\end{center}
\end{figure}

Figure~\ref{current_diff_weak} plots the current as a function of voltage bias for different values of $U$. 
As shown in the right hand panel, the largest reduction of the current is observed for $V/\Gamma\approx 2.5$, which is comparable to the interaction strengths $U/\Gamma=2$, 3.  
As a consistency check, we show in the right hand panel as a thick black curve the interaction correction for $U/\Gamma=2$ deduced from the $4^{th}$ order perturbation calculation of Ref.~[\onlinecite{Fuji03}]. The perfect agreement with the Monte Carlo data shows that the perturbative calculation gives accurate results at this small coupling strength. 

\begin{figure}[t]
\begin{center}
\includegraphics[angle=-90, width=0.49\columnwidth]{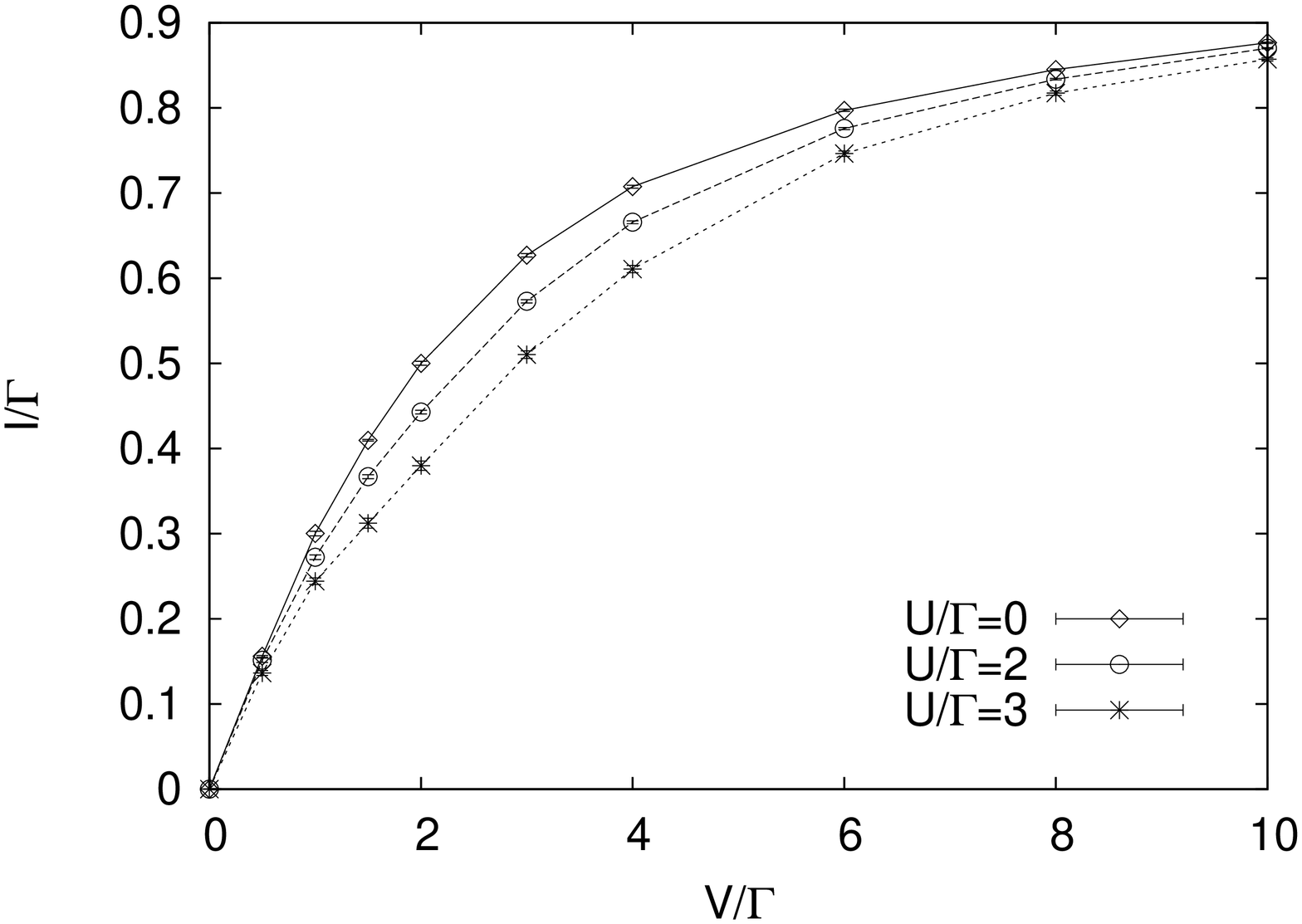}
\includegraphics[angle=-90, width=0.49\columnwidth]{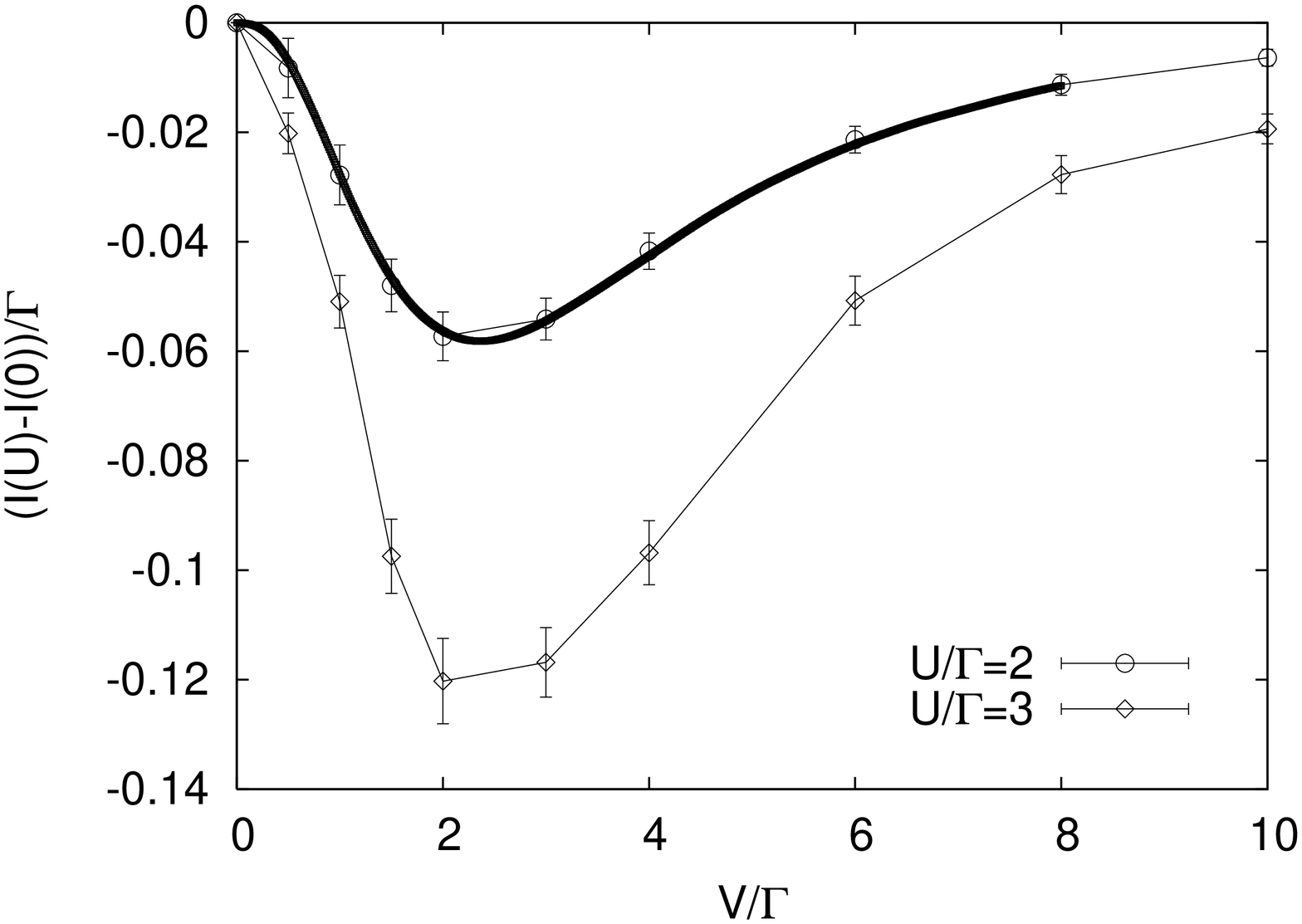}
\caption{
Weak coupling expansion results for the 
voltage dependence of the steady state current for $T=0$ and indicated values of the interaction $U$. The left panel shows the total current and the right panel the reduction in the steady state current produced by the interaction, $I(U)-I(0)$. The interaction correction peaks at a value of $V$ which is comparable to the interaction, or somewhat larger than the twice the ``level broadening" $\Gamma$. The thick black line in the right panel shows the prediction for $U/\Gamma=2$ from $4^{th}$ order perturbation theory.\cite{Fuji03}}
\label{current_diff_weak}
\end{center}
\end{figure}

\subsubsection{Hybridization Expansion}

In the hybridization expansion method (as it has been implemented here) the initial state is an empty dot decoupled from the leads so that at $t=0$ there is no current. As time evolves from $t=0$ the current must build up to its steady state value and the dot occupancy may change. During this transient period, which has been studied in detail in Ref.~[\onlinecite{Schmidt08}], the current into the dot from the right lead ($I_R$) need not equal the current out of the dot into the left lead ($I_L$).  Figure~\ref{current_strong} shows hybridization expansion results for the relaxation dynamics in a dot with voltage bias $V/\Gamma=0$ and 5. 
As the dot-lead hopping is turned on electrons rush from the leads to the initially empty dot, leading to a fast initial rise in the current. For $V=0$ the current from the dot to the left lead ($I_L$) or the right lead ($-I_R$) eventually vanishes. For $V>0$, we see that current initially flows into the dot from both sides, but as time is increased $I_L$ and $-I_R$ converge to equal and opposite non-vanishing  steady state values. The right hand panel shows the difference between the left and right current, which is equal to the derivative of the dot occupation number: $I_L-I_R=dn/dt$, with $n=n_\uparrow+n_\downarrow$. This quantity depends relatively weakly on voltage and converges to zero as the steady state is reached. 

The average current $I=(I_L+I_R)/2$ grows with $V$, as illustrated in the left hand panel of Fig.~\ref{current_u8}. For the parameters in this figure ($U/\Gamma=8$, $\epsilon_d+U/2=0$, band cutoff $\omega_c/\Gamma=10$, $\beta\Gamma=\nu\Gamma=10$) the small oscillations at intermediate times mean that we cannot obtain an accurate estimate of the steady state current. 
In the right hand panel we therefore show the current measured at time $t\Gamma=1$ (solid lines) and 1.25 (dashed lines) as a function of voltage. 
At large voltage, one observes a slow increase of $I(V)$ in the ``Coulomb blockade" regime ($V\lesssim U$) followed by a more rapid increase in the current once the voltage bias exceeds the splitting between the Hubbard bands of approximately $U$. We do not find a rapid increase (comparable to the $U=0$ curve) in the current near $V=0$, presumably because the time-scales reached in this simulation are not long enough for a Kondo resonance to form, or because the latter is destroyed by even a small applied voltage. However, at voltages $V/\Gamma\gtrsim 2$, where the Kondo resonance is wiped out, we expect our hybridization expansion results to be fairly accurate.

Figure~\ref{current_comparison} compares the hybridization expansion results to the mean field and perturbative calculations. The right hand panel shows the comparison to perturbation theory described in Ref.~[\onlinecite{Fuji03}]. We see that for $U\lesssim 6\Gamma$ the results agree quite well.  The deviation seen in the $U=0$ current is due to a difference in bandwidths ($\omega_c/\Gamma=10$ in the Monte Carlo simulation, and infinite bandwidth in the analytical calculation). However, at very small $V$ the hybridization expansion results indicate a lower current than the perturbation expansion of the self-energy. We believe that this difference arises because the hybridization expansion has not been run for long enough times ($\Gamma t=1.25$) to capture the formation of the Kondo (or fermi liquid) resonance. In the perturbative calculation the crossover  from the low $V$ un-renormalized behavior to the larger $V$ suppressed $I(V)$ (visible as a flattening of the perturbative $I(V)$ curve at $V \sim 2\Gamma$ for $U=6\Gamma$) occurs via a voltage-induced splitting of the Kondo resonance, which was also observed in NCA calculations.\cite{Meir93} However, in these calculations the crossover occurs at a voltage far higher than the Kondo temperature, suggesting that the splitting of the Kondo resonance and the associated ``hump" in $I(V)$ might be an artifact and that further investigation of the crossover would be worthwhile. 

\begin{figure}[t]
\begin{center}
\includegraphics[angle=-90, width=0.49\columnwidth]{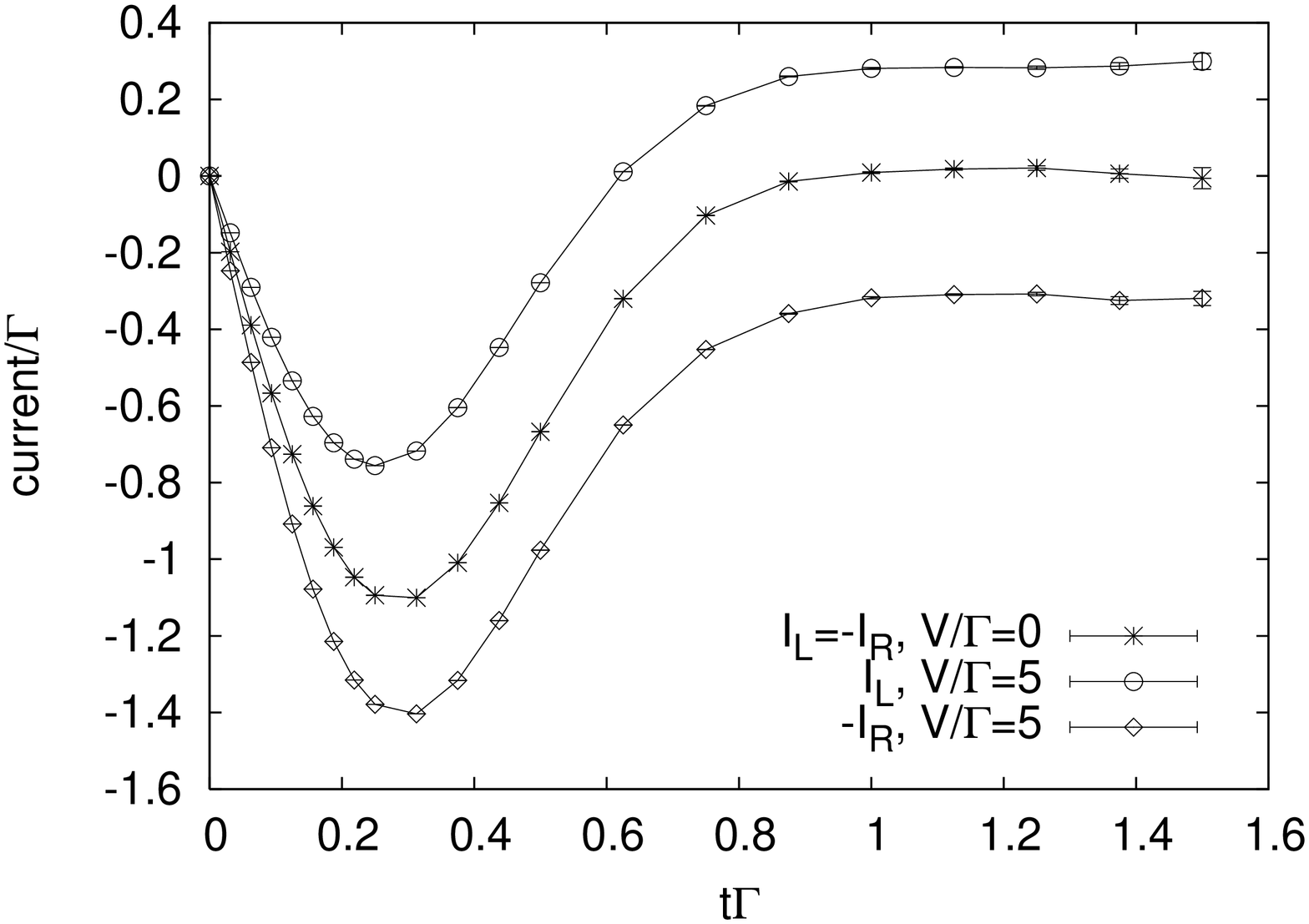}
\includegraphics[angle=-90, width=0.49\columnwidth]{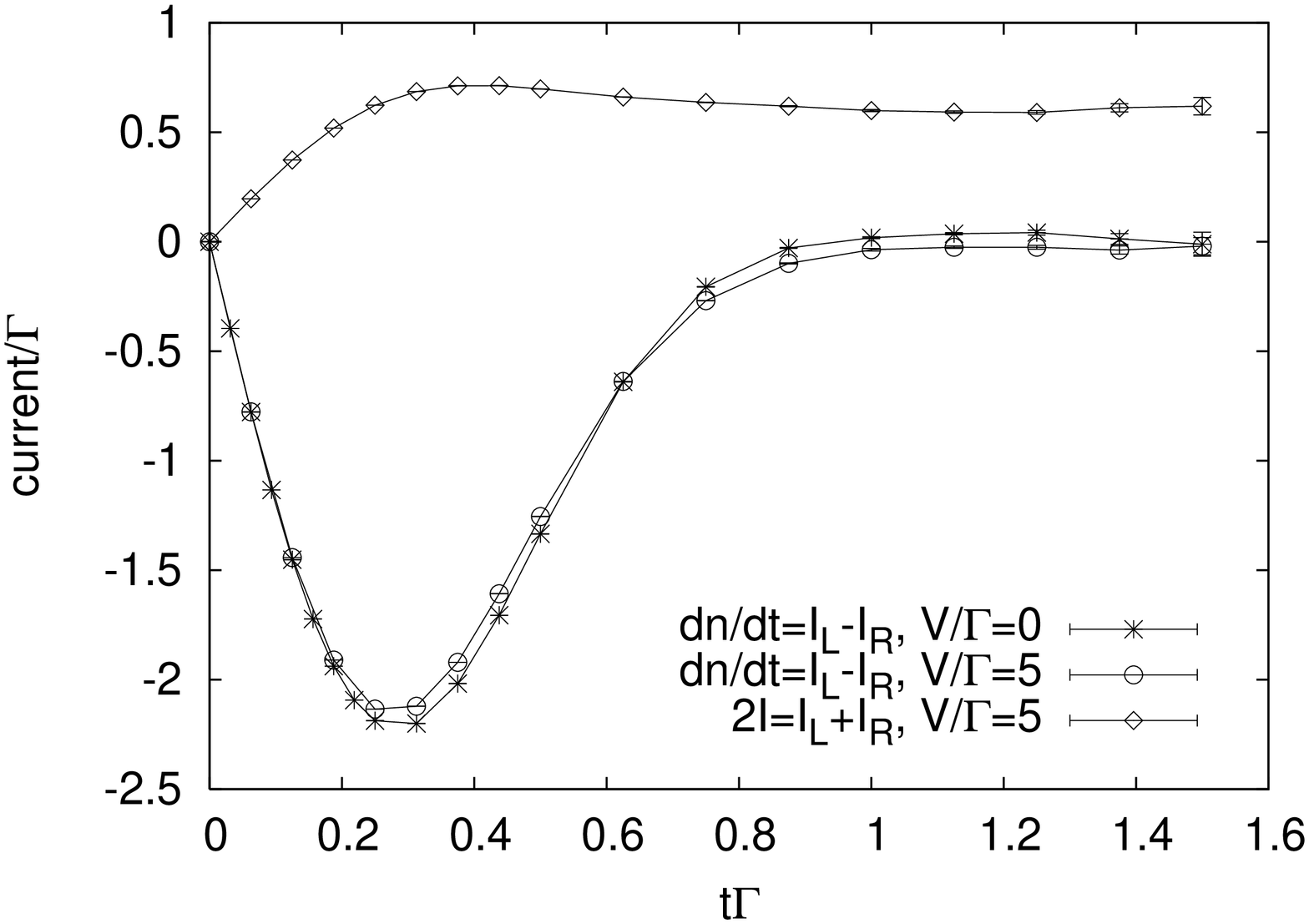}
\caption{Hybridization expansion results for the current through an interacting dot with $U/\Gamma=8$, $\epsilon_d+U/2=0$, and a hard bandcutoff $\omega_c/\Gamma=10$ ($\beta \Gamma = \nu\Gamma=10$). Left panel: left and right current for $V/\Gamma=0$ and $5$. $I_L$ is the current from the dot to the left lead and $I_R$ the current from the right lead to the dot. Right panel: average current $I=(I_L+I_R)/2$ and $dn/dt=I_L-I_R$ for the same parameters.
}
\label{current_strong}
\end{center}
\end{figure}

The left panel of Fig.~\ref{current_comparison} compares the hybridization expansion and mean field calculations. The data for $U=0$ show that the current measured at $t\Gamma=2$ gives a good estimate of the steady state result, especially for larger voltage biases (note that the non-interacting model provides a non-trivial test for the strong coupling method). The mean field theory clearly underestimates the low $V$ current (and in this regime the QMC data probably are themselves an underestimate). However, at larger $V$ the qualitative behavior of the QMC calculations can be more or less reproduced by ``fixed gap" calculations, if   the ``effective temperature" (proportional to $V$) is properly adjusted.  Figure~\ref{current_comparison} also shows that the interacting current approaches the non-interacting value as $V$ becomes very large. The comparison provides evidence of the correctness of the simulation results at large biases. It shows in particular that we are able to access long enough times to obtain reasonable estimates of the asymptotic behavior and suggests that future studies of the pseudothermal broadening effect may be possible. Further investigation of the extent to which Coulomb-blockade-like features persist at high bias and strong coupling would also be of interest.

\begin{figure}[t]
\begin{center}
\includegraphics[angle=-90, width=0.49\columnwidth]{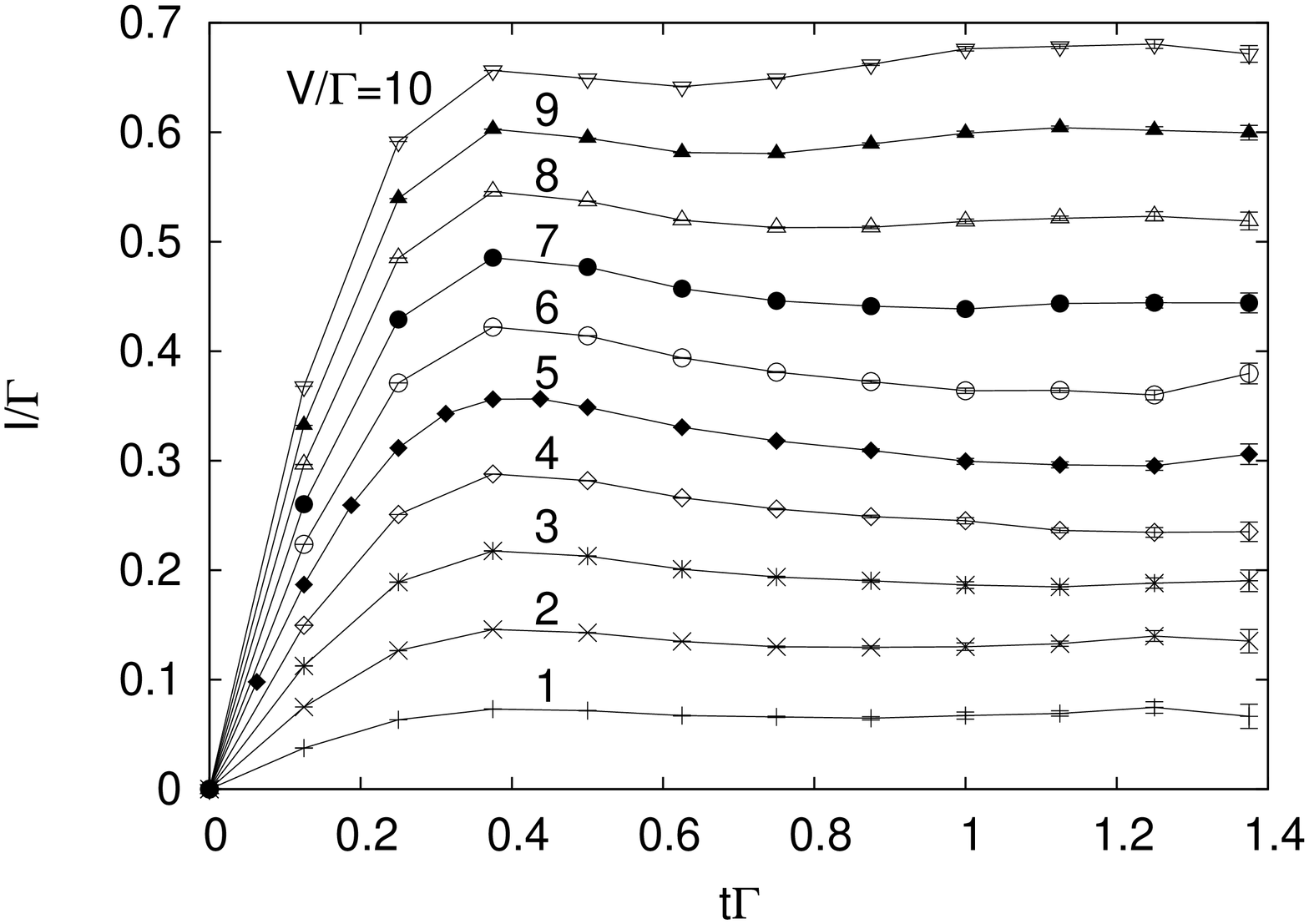}
\includegraphics[angle=-90, width=0.49\columnwidth]{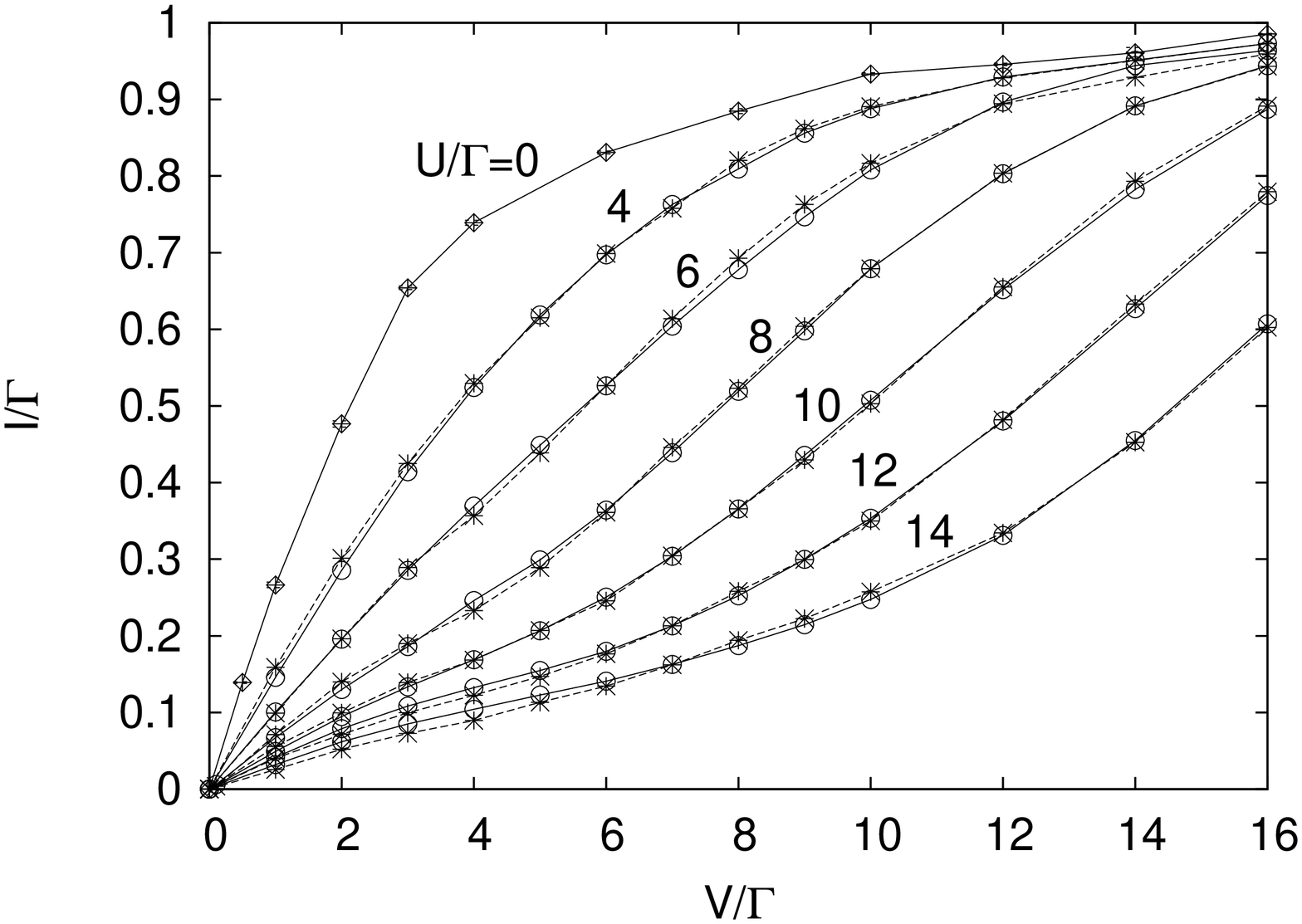}
\caption{Hybridization expansion results for the current through an interacting dot with  
$\epsilon_d+U/2=0$, and a hard bandcutoff $\omega_c/\Gamma=10$ ($\beta \Gamma = \nu\Gamma=10$).  The steady state dot occupancy for this $\epsilon_d$ is 1 (half filling). Left panel: average current $I=(I_L+I_R)/2$ for $U/\Gamma=8$ and indicated values of $V$. Right hand panel: Current at time $t\Gamma=1$ (solid lines) and $t\Gamma=1.25$ (dashed lines) as a function of voltage, for indicated values of the interaction. The current for the non-interacting dot has been measured at $t\Gamma=2$. 
}
\label{current_u8}
\end{center}
\end{figure}

\begin{figure}[t]
\begin{center}
\includegraphics[angle=-90, width=0.49\columnwidth]{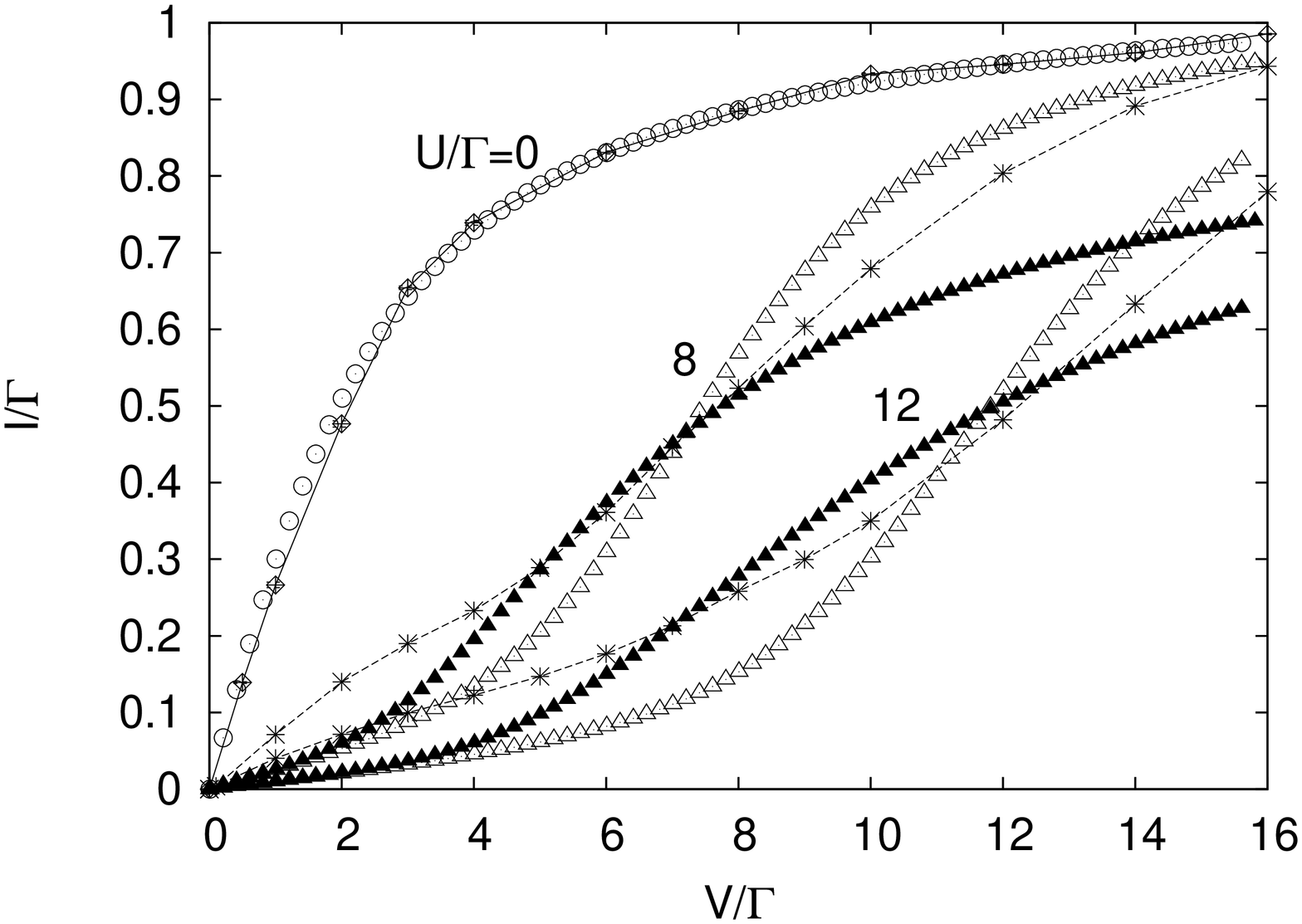}
\includegraphics[angle=-90, width=0.49\columnwidth]{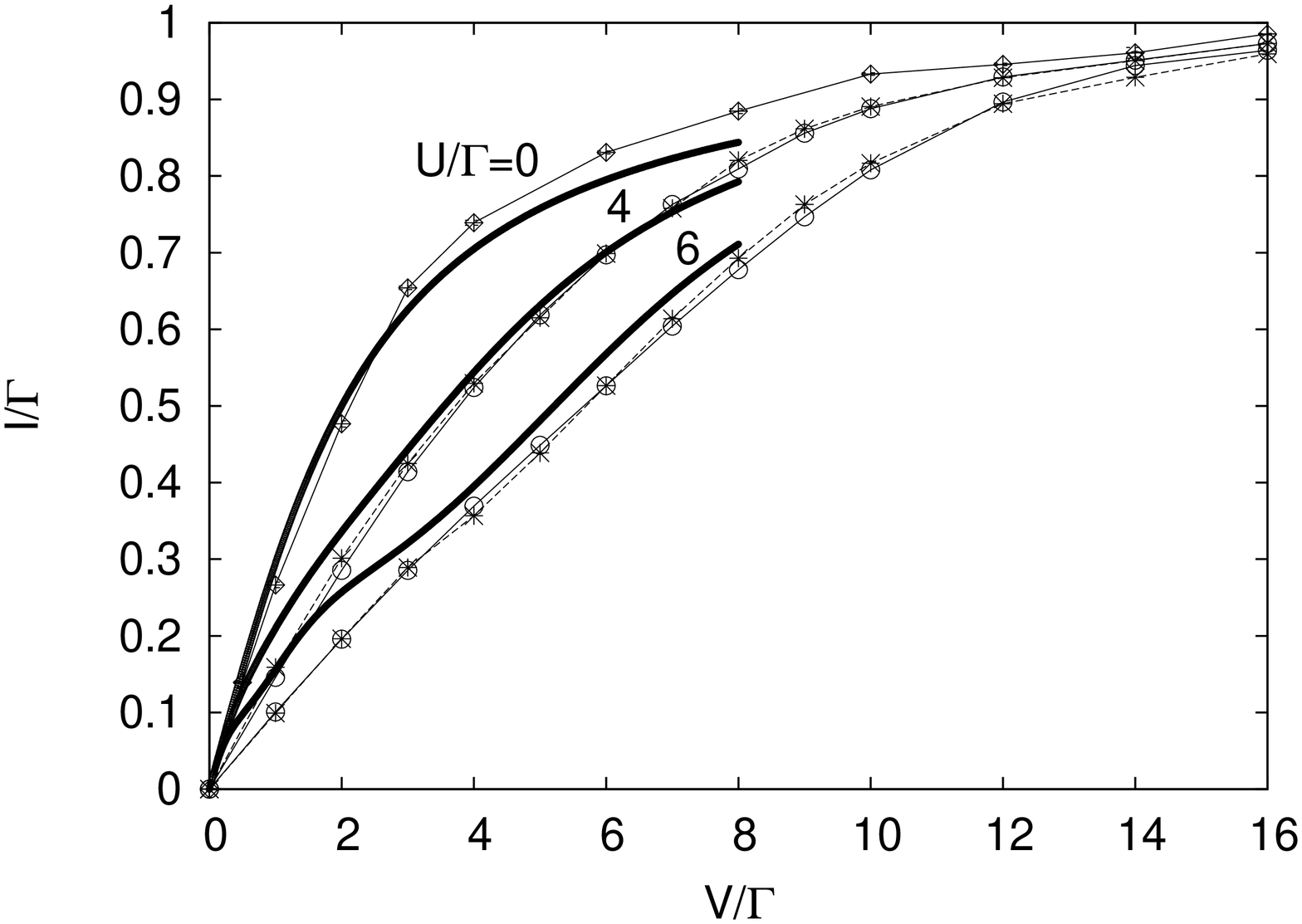}
\caption{Left panel: comparison to the mean-field result. The circles show the (exact) non-interacting current for $\omega_c/\Gamma=10$, $\beta\Gamma=\nu\Gamma=10$, which is in good agreement with the hybridization expansion result measured at $t\Gamma=2$, especially for $V/\Gamma\gtrsim 3$. The triangles show the current obtained with the ``fixed gap" calculation for $U/\Gamma=8$, 12 and an effective temperature $T=0.05V$ (open symbols) and $T=0.2V$ (full symbols). Stars show the Monte Carlo results for $U/\Gamma=8$ and 12 measured at $t\Gamma=1.25$. The right panel compares Monte Carlo results measured at $t\Gamma=1$ (circles), 1.25 (stars) and 2 (diamonds) to the current (for infinite bandwidth) deduced from the $4^{th}$ order perturbation calculation of Ref.~[\onlinecite{Fuji03}].
}
\label{current_comparison}
\end{center}
\end{figure}

\section{Conclusions \label{conclusions}}

In this paper we have investigated an approach to nonequilibrium problems based on a stochastic sampling of diagrams on the Keldysh contour, with diagrams selected on the basis of their contributions to the expectation value being calculated. Both an expansion in interaction strength and an expansion in the dot-lead hybridization were considered. The average expansion order scales linearly with the time interval to be studied.  In both methods the key difficulty is a dynamical sign problem arising from the $i^n$ factors appearing because  one expands $e^{\pm itH}$. The average sign decays exponentially with perturbation order, and when it becomes smaller than about $0.001$ the measurement becomes prohibitively difficult.  

For weakly interacting dots, the weak coupling method 
can be carried to longer times than the hybridization expansion. A further advantage of the weak-coupling method is that it  
starts from an initial density matrix which already contains the entanglement between the dot and the leads, so less time is needed to reach steady state.  However, the growth of the average perturbation order with interaction strength was found to be such that  only interactions in the weak
to intermediate
coupling regime $U \lesssim \pi \Gamma$ can be studied.

The strong coupling method exhibits somewhat worse convergence properties. In interacting dots, only times of the order of $1$-$2$ inverse level widths could be reached. 
A difficulty is that as the method has been formulated here, the initial state is a 
decoupled dot-lead state, which means that the simulation has to build up the necessary dot-lead entanglement before steady state can be reached. 
On the other hand, the method works equally well for all interaction strengths
and the times accessible appear to be long enough that steady state behavior can be reached, at least at large biases where the Kondo effect is not relevant. 
A general advantage of the diagrammatic Monte Carlo technique compared to other methods is that the results are (within the given error bars) exact. There are no discretizations, truncations or other approximations. 
Our results demonstrate that the methods have potential for the simulation of more realistic situations, and may also be able to provide basic insights into issues including the crossover from the Kondo (unrenormalized differential conductance) to high bias regime.

We have not attempted to optimize either of the methods. Better choices of cutoff and of initial conditions are likely to improve the performance of the algorithms. Better sampling procedures, improved estimators or blocking techniques should help reduce the sign problem. Most promising in our opinion are strategies to reduce the average perturbation order, for example through the explicit treatment of bath states in the hybridization expansion approach. Starting the real-time evolution from a thermalized state by sampling configurations on an ``$L$-shaped" contour with an additional branch along the imaginary time direction may lead to a more rapid convergence into the non-equilibrium steady state.
Efforts in these directions are under way. 

\acknowledgements

This research was supported by NSF-DMR-0705847, by a Grant-in-Aid for Young Scientists (B), and the Swiss National Science Foundation (PP002-118866/1). We thank D.~Reichman, L.~M\"uhlbacher, A.~Komnik, and I.~Maruyama for very helpful conversations, and T. Fujii for providing data from Ref.~[\onlinecite{Fuji03}], which we used in Figs.~11, 14 and 17. The calculations were performed on the Hreidar and Brutus clusters at ETH Zurich, using the ALPS library.\cite{alps}

\section{Appendix A: weak coupling formalism \label{AppendixA} }

\subsection{Calculation of $d-d$ Green's function}

We rewrite here for convenience the Hamiltonian for a level coupled to two leads, $\alpha=L,R$ (absorbing the Hartree shift $Un/2$  into the definition of the level energy $\epsilon_d$)
\begin{equation}
H_0=\sum_\sigma\epsilon_dd^\dagger_\sigma d_\sigma +\sum_{k,\sigma,\alpha=L,R}\left(V_k^\alpha c^\dagger_{k,\sigma}d_\sigma+h.c.\right)+\sum_{k,\sigma,\alpha=L,R}\left(\epsilon_k-\mu_\alpha\right)c^\dagger_{k,\sigma,\alpha}c_{k,\sigma,\alpha}.
\label{hdef}
\end{equation}

The physics associated with the  coupling  to the leads can be reconstructed from
\begin{equation}
\Gamma_\alpha(\omega)=\pi\sum_k \left|V^\alpha_k\right|^2\delta(\omega-\epsilon_k)
\label{gamdef}
\end{equation}
and
\begin{equation}
S_\alpha(\omega)=\int \frac{dx}{\pi}{\cal P}\frac{\Gamma_\alpha(x)}{\omega-x}
\end{equation}
with ${\cal P}$ the principal value symbol. In the infinite bandwidth, constant density of states limit $\Gamma_\alpha$ is constant and $S_\alpha(\omega)=0$.

The coupling to the leads provides a self energy  $\Delta$ to the dot Green's function. Because the leads are infinite the self energy may be computed in terms of the correlators $G_\text{cond}$ of the $c$ electrons with hybridization $V=0$.\cite{Mahan} In the Larkin basis the calculation follows the same lines as the equilibrium \cite{Mahan} one with
\begin{equation}
\Delta^{R,A,K}_\alpha=\sum_k\left|V_{k}^\alpha\right|^2G^{R,A,K}_{\text{cond},\alpha}(k,\omega),
\label{delta}
\end{equation}  
where ($\delta$ is the usual positive infinitesimal and the upper (lower) sign pertains to $G^R$ ($G^A$))
\begin{eqnarray}
G^{R/A}_{\text{cond},\alpha}(k,\omega)&=&\frac{1}{\omega-\varepsilon_k\pm i\delta},
\label{gcondra} \\
G^K_{\text{cond},\alpha}(k,\omega)&=&-2\pi i\delta\left(\omega-\varepsilon_k\right)\tanh\left(\frac{\omega-\mu_\alpha}{2T_\alpha}\right).
\label{gcondk}
\end{eqnarray}

Inserting Eqs.~(\ref{gcondra}), (\ref{gcondk}) into Eq.~(\ref{delta}) gives
\begin{eqnarray}
\Delta^R_\alpha(\omega)&=&S_\alpha(\omega)-i\Gamma_\alpha(\omega),
\\
\Delta^A_\alpha(\omega)&=&S_\alpha(\omega)+i\Gamma_\alpha(\omega),
\\
\Delta^K_\alpha(\omega)&=&-2i\Gamma_\alpha(\omega)\tanh\left(\frac{\omega-\mu_\alpha}{2T_\alpha}\right).
\end{eqnarray}

Then using the symbol without the $\alpha$ subscript to denote the sum of left and right channel contributions (so e.g. $\Delta^{R,A,K}=\Delta^{R,A,K}_L+\Delta^{R,A,K}_R$ etc.) we find that  the full $d$ Green's function ${\bf G}_{dd}$ is given by
\begin{equation}
\left(\begin{array}{cc}G_{dd}^R & G_{dd}^K \\0 & G_{dd}^A\end{array}\right)=\left(\left(\begin{array}{cc}\omega-\varepsilon_d & 0 \\0 & \omega-\varepsilon_d\end{array}\right)-\left(\begin{array}{cc}\Delta^R(\omega) & \Delta^K(\omega) \\0 & \Delta^A(\omega)\end{array}\right)\right)^{-1}.
\label {Gdd}
\end{equation}

Use of the standard relations \cite{Rammer96} gives 
\begin{eqnarray}
G^>_{dd}&=&\frac{1}{2}\left(G^K_{dd}+G^R_{dd}-G^A_{dd}\right)=\sum_\alpha\frac{-i\Gamma_\alpha(1+\tanh((\omega-\mu_\alpha)/(2T_\alpha)))}{(\omega-\epsilon_d-S)^2+\Gamma^2},\label{G>def}\\
G^<_{dd}&=&\frac{1}{2}\left(G^K_{dd}-G^R_{dd}+G^A_{dd}\right)=\sum_\alpha\frac{i\Gamma_\alpha(1-\tanh((\omega-\mu_\alpha)/(2T_\alpha)))}{(\omega-\epsilon_d-S)^2+\Gamma^2}\label{G<def}.
\end{eqnarray}

\subsection{calculation of $A(t,t')$}

We  express the quantity $A(t,t')=\langle \tilde a^\dagger_{L}(t') d(t) \rangle_0$  (with retarded/advanced/Keldysh nature here left unspecified) as
\begin{equation}
{\bf A}=-i\sum_kV^L_kG^{cd}_k.
\label{Adef}
\end{equation}
Use of the equation of motion that led to Eq.~(11) of  Ref.~[\onlinecite{Jauho94}]  gives
(denoting convolution by products) 
\begin{eqnarray}
\left(\begin{array}{cc}A^R & A^K \\0 & A^A\end{array}\right)
&=&-i\left(\begin{array}{cc}G_{dd}^R & G_{dd}^K \\0 & G_{dd}^A\end{array}\right)\left(\begin{array}{cc}\Delta_L^R & \Delta_L^K \\0 & \Delta_L^A\end{array}\right)
=-i\left(\begin{array}{cc}G_{dd}^R\Delta_L^R & G_{dd}^R\Delta_L^K+G_{dd}^K \Delta_L^A\\0 & G_{dd}^A\Delta_L^A\end{array}\right).
\label{Gcd1}
\end{eqnarray}
From Eqs.~(\ref{Gdd}) and (\ref{Gcd1}) we find 
\begin{eqnarray}
A^R&=&-i\frac{\left(\omega-\varepsilon_d-S-i\Gamma\right)\left(S_L-i\Gamma_L\right)}{(\omega-\varepsilon_d-S)^2+\Gamma^2}, \\
A^A&=&-i\frac{\left(\omega-\varepsilon_d-S+i\Gamma\right)\left(S_L+i\Gamma_L\right)}{(\omega-\varepsilon_d-S)^2+\Gamma^2}, \\
A^K&=&-i\frac{\left(\omega-\varepsilon_d-S-i\Gamma\right)(-2i\Gamma_Lh_L)-2i(\Gamma_Lh_L+\Gamma_Rh_R)\left(S_L+i\Gamma_L\right)}{(\omega-\varepsilon_d-S)^2+\Gamma^2}
\nonumber \\
&=&\frac{(-2\Gamma_Lh_L)\left(\omega-\varepsilon_d\right)+2\Gamma_Lh_L\left(S_R+i\Gamma_R\right)-2\Gamma_Rh_R\left(S_L+i\Gamma_L\right)}{(\omega-\varepsilon_d-S)^2+\Gamma^2},
\label{gcdk}
\end{eqnarray}
with $h_\alpha=\tanh((\omega-\mu_\alpha)/(2T_\alpha))$ and $\Gamma=\Gamma_L+\Gamma_R$. 
Therefore, with $f_\alpha\equiv f(\omega-\mu_\alpha)$ denoting the Fermi function for lead $\alpha$ and use of the relation $A^<=\frac{1}{2}(A^K-A^R+A^A)$ we find
\begin{eqnarray}
A^<&=&-2\frac{i\Gamma_L\Gamma_R\left(f_L-f_R\right)-\Gamma_Lf_L\left(\omega-\varepsilon_d\right)+S_R\Gamma_Lf_L-S_L\Gamma_Rf_R}{(\omega-\varepsilon_d-S)^2+\Gamma^2}.
\label{gdcf>inal}
\end{eqnarray}
Setting $S=0$ gives
\begin{eqnarray}
A^<&=&-2\frac{i\Gamma_L\Gamma_R\left(f_L-f_R\right)-\Gamma_Lf_L\left(\omega-\varepsilon_d\right)}{(\omega-\varepsilon_d)^2+\Gamma^2}.
\label{gdcf>simple}
\end{eqnarray}

\subsection{Mean Field Theory}

In the mean field theory \cite{Komnik04} of the nonequilibrium Anderson model one replaces the Hamiltonian by 
\begin{eqnarray}
H_{MF}=\sum_\sigma\varepsilon_\sigma d^\dagger_\sigma d_\sigma+\sum_{\alpha,p,\sigma}\left(V_{\alpha,p} d^\dagger_\sigma c_{\alpha, p, \sigma}+V_{\alpha,p}^\star c^\dagger_{\alpha, p, \sigma} d_\sigma\right)+\sum_{\alpha, p, \sigma}\varepsilon_p c^\dagger_{\alpha, p, \sigma}c_{\alpha, p,\sigma}
\end{eqnarray}
with
\begin{equation}
\varepsilon_\sigma=\varepsilon_0+Un_{-\sigma}.
\label{sce1}
\end{equation}
The occupancy of the $d$-orbital of spin $\sigma$ is then 
\begin{eqnarray}
n_{d\sigma}=\int \frac{d\omega}{\pi}\frac{\Gamma_L(\omega)f(\omega-\mu_L)+\Gamma_Rf(\omega-\mu_R)}{(\omega-\varepsilon_\sigma-S(\omega))^2+\Gamma(\omega)^2}
\label{sce2}
\end{eqnarray}
and one requires self-consistency between Eqs.~(\ref{sce1}) and (\ref{sce2}). In practice self consistency is achieved by starting from an initial guess and iterating until the equations cease to change. 

In our explicit calculations we took a flat band with a hard cutoff defined by 
\begin{eqnarray}
\Gamma_{L,R}(\omega)&=&0.5\Gamma\left(\tan^{-1}\left[\frac{\omega_c+\omega}{\delta}\right]+
\tan^{-1}\left[\frac{\omega_c-\omega}{\delta}\right]\right),
\\
S_{L,R}&=&\frac{0.5\Gamma}{2\pi}\ln\left[\frac{(\omega_c-\omega)^2+\delta^2}{(\omega_c+\omega)^2+\delta^2}\right],
\end{eqnarray}
with $\omega_c=10\Gamma$ and $\delta=0.1\Gamma$.

We choose conventions such that $\mu_L+\mu_R=0$ and $\mu_L-\mu_R=V$. Then the current is given by 
\begin{equation}
I=\sum_\sigma\int\frac{d\omega}{2\pi}\frac{2\Gamma_L(\omega)\Gamma_R(\omega)\left(f(\omega-\mu_L)-f(\omega-\mu_R)\right)}{(\omega-\varepsilon_\sigma-S(\omega))^2+\Gamma(\omega)^2}.
\end{equation}
To represent the broadening effect of a voltage bias, in some calculations we include in the Fermi functions an effective temperature equal to a constant times the voltage bias.

\end{document}